\documentclass[fleqn,usenatbib]{mnras}
\usepackage{newtxtext,newtxmath}
\usepackage[T1]{fontenc}
\DeclareRobustCommand{\VAN}[3]{#2}
\let\VANthebibliography\thebibliography
\def\thebibliography{\DeclareRobustCommand{\VAN}[3]{##3}\VANthebibliography}

\usepackage{graphicx}	
\usepackage{amsmath}	
\usepackage{physics}

\title[Non-parametric debris disk structure]{Recovering the structure of debris disks non-parametrically from images}

\author[Y. Han et al.]{
Yinuo Han,$^{1,2}$\thanks{E-mail: yinuo@caltech.edu}
Mark~C.~Wyatt$^{2}$ and
Sebastian~Marino$^{3}$
\\
$^{1}$Division of Geological and Planetary Sciences, California Institute of Technology, 1200 E. California Blvd., 91125, Pasadena, CA, USA\\
$^{2}$Institute of Astronomy, University of Cambridge, Madingley Road, Cambridge CB3 0HA, UK\\
$^{3}$Department of Physics and Astronomy, University of Exeter, Stocker Road, Exeter, EX4 4QL, UK\\
}

\date{Accepted XXX. Received YYY; in original form ZZZ}

\pubyear{2024}

\begin{document}
\label{firstpage}
\pagerange{\pageref{firstpage}--\pageref{lastpage}}
\maketitle

\begin{abstract}
Debris disks common around Sun-like stars carry dynamical imprints in their structure that are key to understanding the formation and evolution history of planetary systems. 
In this paper, we extend an algorithm (\texttt{rave}) originally developed to model edge-on disks to be applicable to disks at all inclinations. The updated algorithm allows for non-parametric recovery of the underlying (i.e., deconvolved) radial profile and vertical height of optically thin, axisymmetric disks imaged in either thermal emission or scattered light. Application to simulated images demonstrates that the de-projection and deconvolution performance allows for accurate recovery of features comparable to or larger than the beam or PSF size, with realistic uncertainties that are independent of model assumptions. We apply our method to recover the radial profile and vertical height of a sample of 18 inclined debris disks observed with ALMA. Our recovered structures largely agree with those fitted with an alternative visibility-space de-projection and deconvolution method (\texttt{frank}). We find that for disks in the sample with a well-defined main belt, the belt radius, fractional width and fractional outer edge width all tend to increase with age, but do not correlate in a clear or monotonic way with dust mass or stellar temperature. In contrast, the scale height aspect ratio does not strongly correlate with age, but broadly increases with stellar temperature. 
These trends could reflect a combination of intrinsic collisional evolution in the disk and the interaction of perturbing planets with the disk’s own gravity. 
\end{abstract}

 

\section{Introduction}
\label{sec:introduction}

Planets are thought to form in dusty, gas-rich disks, in which protoplanets accrete mass until the dispersal of the protoplanetary disk \citep{Williams2011, Andrews2020}. However, a large population of planetesimals formed in this process never go on to become part of a planet. They become minor bodies in the newly formed planetary system, constituting a debris disk \citep{Wyatt2008, Matthews2014}. The asteroid belt and Kuiper belt are such examples in the Solar System. 

Although individual planetesimals in debris disks are too small to be directly observed, constant collisions between them feed a ``collisional cascade'' in which bodies grind down into successively smaller grains until the smallest grains are removed from the system by radiation pressure \citep{Tanaka1996}. These dust grains collectively contain a large surface area, and are the signatures of solid bodies in debris disks which we are able to observe. It is estimated that approximately 20\% of Sun-like (i.e., F-, G- or K-type) stars host detectable levels of dust \citep{Sibthorpe2018}, many of which have been imaged via either light scattered from the star in optical to near-infrared wavelengths or thermal emission in mid-infrared to millimeter wavelengths \citep{Hughes2018}. 

The debris disks which we can resolve with current instruments are typically located at 10s of au or above, making them analogues of the Solar System’s Kuiper belt in the outer planetary system, rather than the asteroid belt in the inner regions. Owing to the difficulty in observing outer planets, debris disks often provide the best constraints on the outer planetary system. 

Obtaining these constraints on the architecture and evolutionary history of planetary systems relies on a robust understanding of the disk structure. Imaging campaigns in recent decades have produced a repository of dozens of resolved disks from optical to millimeter wavelengths. Sub-structures have been found to be common in debris disks, motivating a host of theoretical work to explain the formation of these structures. Notable examples include gaps potentially carved by planets (e.g., in HR8799, \citealp{Faramaz2021}, HD92945, \citealp{Marino2018}, and HD107146, \citealp{Marino2019}), clumps possibly caused by resonant trapping of planetesimals or giant impacts (e.g., in $\epsilon$~Eri, \citealp{Booth2023}, $\beta$~Pic, \citealp{Telesco2005}), warped (e.g., in $\beta$~Pic, \citealp{Mouillet1997}) or eccentric (e.g., in Fomalhaut, \citealp{MacGregor2017Fom}) disks thought to be caused by secular perturbation from inclined planets and smoothly decaying disk outer edges potentially caused by migrating planets scattering material into high-eccentricity orbits (e.g, in q$^1$~Eri, \citealp{Lovell2021}).

As current and future instruments such as JWST, ALMA and the ELT continue to improve our imaging capabilities, the number of resolved disks and the quality of images are expected to be further increased. As part of the effort to understand debris disk structures from these images, methods to systematically understand their three-dimensional structures in an unbiased way will be critical. 

The foremost task in recovering the radial disk structure from an image is to de-project inclination effects in the presence of a vertical thickness and de-convolve point spread function (PSF) or synthesized beam effects. Due to the effect of convolution with the PSF/beam, each point in the physical disk contributes flux to all locations in the disk image. Furthermore, for edge-on disks, each projected location along the disk receives flux contribution from a range of radii which needs to be disentangled to recover the radial profile. Studies in the literature commonly take a parametric approach to address this, assuming a functional form for the radial distribution of the surface brightness and vertical height and using an automated sampling algorithm such as a Markov chain Monte Carlo (MCMC) to optimise the functional parameters (e.g., \citealp{Marino2021}). Although the disk structure is often broadly visible from the image, a quantitative understanding of features such as slopes and widths is important and affect theoretical interpretation and constraints on the planetary system. Unbiased methods are therefore important. While a parametric approach is able to constrain parameters of interest under an assumed geometric model or dynamical scenario, the fitted radial profiles are inevitably biased to comply with the model assumed in the first place, and the uncertainties are conditioned upon the assumed model being ground truth, causing them to be underestimated.

Ideally, we would be able to achieve an effective ``non-parametric'' structural recovery method, which would avoid requiring assumptions on the functional form when performing de-projection and de-convolution while providing uncertainties that are largely model-independent.
In the context of debris disks, such approaches have already been worked on. \citet{Han2022} presented the \texttt{rave} algorithm that non-parametrically recovers the radial profile independent of knowledge on the vertical height and inclination of the disk. For sufficiently edge-on disks, the algorithm can also recover the scale height as a function of radius non-parametrically. As \texttt{rave} operates in image space by fitting discrete annuli, it is applicable to images at all wavelengths with isotropic emission (e.g., thermal emission in the mid-infrared or millimeter) as long as the disk is azimuthally symmetric and optically thin along the line of sight at the observed wavelength. \citet{Jennings2020} also presented the \texttt{frank} de-projection and deconvolution algorithm application to interferometric observations (such as with ALMA), which recovers the radial profile directly from visiblity data assuming axisymmetry. 

While the broad applicability of \texttt{rave} would in theory provide a method for recovering disk structures across wavelength, imaging mode and inclination, the present algorithm one-dimensionalises the image along the major axis without making use of the 2D information available for disks at low inclination (i.e., closer to face-on). The method could thus ideally be improved and optimised for more face-on systems. Furthermore, \texttt{rave} presently assumes isotropic emission, but it may be possible to incorporate scattering phase function effects to model scattered light observations, in which the angle between incident starlight and the line of sight varies with azimuth across the disk. 

In this paper, we present face-on and scattered light extensions to the \texttt{rave} method. This in theory extends the applicability of \texttt{rave} to disk observations at all wavelengths and optimized for all inclinations, offering a method to uniformly study the structure in resolved disk images. We describe the algorithm in Section~\ref{sec:algorithm} and assess its performance with test cases in Section~\ref{sec:demonstration}. We then apply the method to recover the radial profile and vertical height of a sample of ALMA-observed debris disks in Section~\ref{sec:application} and discuss the disk and stellar properties in the sample. The findings are summarised in Section~\ref{sec:conclusions}.

\section{Algorithm}
\label{sec:algorithm}
In this section, we describe two extensions to the \texttt{rave} algorithm presented in \citet{Han2022} that optimises and generalises its applicability. The first optimises the fitting procedure for disks that are sufficiently face-on to measure meaningful azimuthally averaged radial profiles, and the second enables fitting to scattered light observations by taking into account the effect of the scattering phase function. 

\subsection{Optimisation for face-on disks}
\label{sec:fave}

\begin{figure*}
    \centering
    \includegraphics[width=18cm]{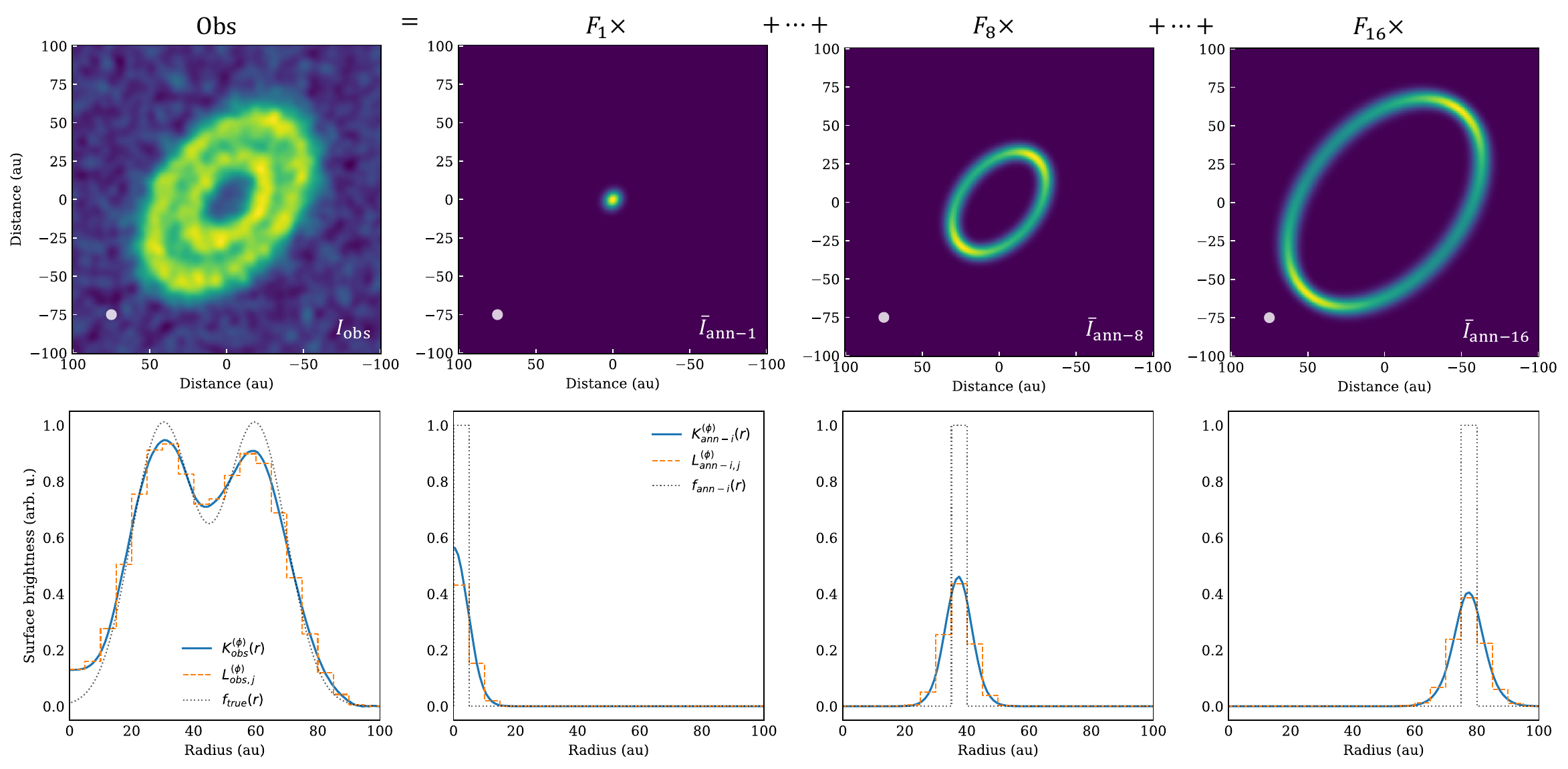}
    \caption{Diagram showing the decomposition of a disk (left column) into constituent annuli (3 columns on the right). The top row shows the observed images and simulated (and convolved) images of the annuli, and the bottom row shows their azimuthally averaged profiles. }
    \label{fig:fave}
\end{figure*}

\subsubsection{Radial surface brightness profile}
\label{sec:fave-radial}
Prior to this work, the \texttt{rave} non-parametric fitting method was already applicable to disks viewed at any inclination, but here we present improvements to the algorithm for non-edge-on disks. 

To summarise the present algorithm, we largely follow the notation used in \citet{Han2022} with minor modifications. Assuming the disk to be optically thin and azimuthally symmetric, the algorithm treats the disk image being fitted to, $I_\text{obs}$, as the weighted sum of a series of $N$ neighbouring concentric annuli,
$\bar I_{\text{ann-} i}$, boundaries $\vb*{R}=(0, R_1, \dots, R_N)$, vertical height $\vb*{H}=(H_1, \dots, H_N)$ and inclination $\iota$, and the images are convolved with the same PSF/beam as in the observations being fitted to. 
The relationship between the disk image and its constituent annuli, each with unity surface brightness, can be written as
\begin{equation}
    I_\text{model} = \sum_{i=1}^{N} F_i \, \bar I_{\text{ann-}i}.
    \label{eq:annuli}
\end{equation}
The primary task is then to solve for the weighting (i.e., brightness) required of each individual annulus, $\vb* F = (F_1, F_2, ..., F_{N})$, that reproduces the observed disk image.

To recover the de-projected and deconvolved surface brightness profile, $\vb* F$, the approach first involves reducing each 2D image of the disk and normalised annuli from $I$ into a characteristic 1D quantity, $K$, which corresponds to the emission of the disk summed onto the major axis, 
\begin{equation}
    K_\text{obs}^{(y)}(x) = \int_{-\infty}^{\infty} I_\text{obs}(x, y)dy,
\end{equation}
\begin{equation}
    \bar K_{\text{ann-}i}^{(y)}(x) = \int_{-\infty}^{\infty} \bar I_{\text{ann-}i}(x, y, i, H_i)dy,
    \label{eq:ky}
\end{equation}
\noindent where $x$ and $y$ are the projected displacements from the star along the major and minor axis respectively. Note that the dependence on the vertical height and inclination assumed when simulating annuli images, $\bar I_{\text{ann-}i}$, vanishes when reducing the image into the one-dimensional flux, $\bar K_{\text{ann-}i}^{(y)}$. One-dimensionalising the problem implies that the surface brightness could then be solved for via a system of $N$ linear equations, with each equation describing the flux contribution from all annuli to a region, $R_{j-1} < x \leq R_j$, along the disk. Details of this algorithm are described in Section~2.1 in \citep{Han2022}. 

Having summed over the spatial distribution along the minor axis, $K_\text{obs}^{(y)}$ is independent of the inclination and vertical height of the disk. The fitting procedure is therefore applicable to disks at all inclinations and the recovered radial profile is independent of knowledge on the inclination and vertical height. While such an approach is convenient for edge-on disks, it also discards 2D information along the minor axis which is in fact useful for more face-on disks, in which the large-scale radial structure is already visible from the image, albeit PSF/beam-convolved and projected. So while the algorithm is applicable for disks at all inclinations, we aim to optimise the de-projection and deconvolution for non-edge-on disks to make use of the additional information and achieve a more accurate radial profile recovery. 

To achieve this, we replace the vertically summed flux, $K^{(y)}$, in non-edge-on disks with the azimuthally averaged flux profile,
\begin{equation}
    K_\text{obs}^{(\phi)}(r, \iota) = \frac{1}{2\pi}\int_{0}^{2\pi} I_\text{obs}(r, \phi)d\phi,
    \label{eq:kphi-obs}
\end{equation}
\begin{equation}
    \bar K_{\text{ann-}i}^{(\phi)}(r, \iota, H_i) =\frac{1}{2\pi}\int_{0}^{2\pi} \bar I_{\text{ann-}i}(r, \phi, \iota, H_i)d\phi,
    \label{eq:kphi}
\end{equation}
\noindent where we use $(r, \phi)$ to denote polar coordinates in the plane of the disk and $(r', \phi')$ for coordinates in the projected image plane, and $\iota$ is the inclination of the disk. 
Fig.~\ref{fig:fave} shows examples of $K$ and other relevant quantities.

Note that $\iota$ dependence of $K^{(\phi)}$ arises from the conversion between $(r, \phi)$ and $(r', \phi')$, given by $r' = r \sqrt{\cos{\phi}^2 + \sin{\phi}^2\cos{\iota}^2}$ and $\cos(\phi') = 1 / \sqrt{1 + \tan{\phi}^2 \cos{\iota}^2}$. For $\bar K_{\text{ann-}i}^{(\phi)}$, the $\iota$ dependence also arises from the inclination assumed when simulating normalised annuli images, $\bar I_{\text{ann-}i}$. For well-resolved disks in which \texttt{rave} can be meaningfully applied, the inclination can generally be readily constrained from the large-scale disk geometry observed. We thus omit the explicit $\iota$ dependence in subsequent equations, assuming that $\iota$ is well-constrained from the observations and that all annuli are simulated with the observationally constrained inclination. 

Substituting $K$ for $I$ in Eq.~\eqref{eq:annuli}, we obtain
\begin{equation}
    K_\text{model}^{(\phi)}(r) = \sum_{i=1}^{N} F_i \, \bar K_{\text{ann-}i}^{(\phi)}(r, H_i).
    \label{eq:k_annuli}
\end{equation}
To obtain an exact solution for the $N$ components of $\vb* F$, $N$ linearly independent equations are required. We partition $r$ into $N$ discrete neighbouring bins, with each bin corresponding to the projected location of the constituent annuli defined by the boundaries $\vb* R$. This in effect discretises $K_\text{obs}^{(\phi)}(r)$ and $K_\text{model}^{(\phi)}(r)$ into $N$ components, each defined by
\begin{equation}
    L^{(\phi)}_j = \frac{1}{R_j - R_{j-1}} \int_{R_{j-1}}^{R_j} K^{(\phi)}(r) dr,
\end{equation}
which in turn results in $N$ equations derived from Eq.~\eqref{eq:k_annuli}, each given by
\begin{equation}
     L_{\text{model}, j}^{(\phi)} = \sum_{i=1}^{N} F_i \, \bar L_{\text{ann-}i, j}^{(\phi)}(H_i).
    \label{eq:l_annuli}
\end{equation}
Eq.~\eqref{eq:l_annuli} can be equivalently written in matrix form, 
\begin{equation}
    \vb*{L}_{\text{model}}^{(\phi)} = M^{(\phi)}(H_i) \ \vb*F,
\end{equation}
where $M^{(\phi)}_{j, i}(H_i) = \bar L_{\text{ann-}i, j}^{(\phi)}(H_i)$. The de-projected and deconvolved radial profile, $\vb* F$, can then be solved for by setting $\vb*{L}_{\text{model}}^{(\phi)} = \vb*{L}_{\text{obs}}^{(\phi)}$ and inverting $M$:
\begin{equation}
    \vb*F = M^{(\phi)-1}(H_i) \ \vb*{L}_{\text{obs}}^{(\phi)}.
    \label{eq:matrix_inversion}
\end{equation}

A few remaining details of the algorithm are identical to those in the original formulation for edge-on disks. Firstly, the annuli images are generated using a Monte Carlo simulation by drawing random points that are uniformly distributed within the disk plane and normally distributed in the vertical direction with a scale height of $H_i$. Details on simulating annuli images are described in Section~2.1.2 in \citet{Han2022}. Secondly, the updated algorithm can still fit to the stellar flux by including an additional annulus in the fit with a radius and height of 0, which effectively acts as a point source. Thirdly, the procedure thus far recovers a version of the radial profile that is discrete (i.e., piecewise constant) without uncertainties and the result is dependant on the location of annuli boundaries defined when fitting. To obtain a continuous radial profile with uncertainties that is robust against the choice of annuli location, we repeat the fit a large number of times ($>$100), each time adopting a randomised set of annuli boundary locations. The median among all the fitted discrete radial profiles produces a smooth median model, $f(r \mid H(r))$, analogous to the ``best-fit model'' in parametric fitting, and the standard deviation among them defines the ``range of possible models'', which differs from uncertainties in parametric fitting in that it encapsulates a range of different models with varying functional forms. Details of this Monte-Carlo fitting approach are described in Section~2.1.4 in \citet{Han2022}.

\subsubsection{Constraining the vertical height}
\label{sec:fave-vertical}

The procedure described in Section~\ref{sec:fave-radial} recovers the de-projected and deconvolved radial profile $f(r \mid H(r))$ given a height assumption. In this section, we develop a method to constrain the vertical height, which in turn motivates the height assumption when deriving the best-fit radial profile. 

While replacing $K^{(y)}$ with $K^{(\phi)}$ for non-edge-on disks improves the robustness of the fit against noise as Section~\ref{sec:demonstration} will demonstrate, this does imply that the fitted radial profile depends on the height assumption, unlike in the edge-on algorithm. However, the same dependence could in fact be leveraged to constrain the vertical height, which relies on the realisation that the fitted model best reproduces the observations (as measured by squared residuals between the observed and modelled images) when the height assumption is equal to the true height. Such a correspondence is expected since the algorithm only fits to the azimuthally averaged profile, which is determined by both the radial surface brightness and height profiles in a degenerate way, but the two have different effects on the 2D image, which theoretically could only be reproduced if the height assumption is equal to the true height of the disk. 

To simplify the problem, we assume that the vertical height is proportional to radius, such that $H(r) = hr$, where $h$ is the scale height aspect ratio. Suppose that under an aspect ratio assumption, $h$, the recovered radial profile is $f(r \mid h)$. We may define the normalised squared residuals as
\begin{equation}
    \chi^2(h) = \frac{\nu}{\sigma_b^2 \Omega_I}\int_x \int_y [I_\text{obs}(x, y) - I_\text{model}(x, y, f(r \mid h), h)]^2 dx dy,
\end{equation}
where $\sigma_b$ is the Root-Mean-Square (RMS) noise per beam, $\Omega_I$ is the area within the field of view of the image and $\nu$ is the number of independent elements in the image of the disk (approximated with the number of beams for an ALMA image). Assuming that the noise is independently and identically Gaussian distributed, the likelihood of a given $h$ given the observations and the recovered radial profile is then
\begin{equation}
    P(h) \propto e^{-\chi^2/2}.
\end{equation}

In practice, we sample $P(h)$ by generating a grid of $h$ values, obtaining a de-projected and deconvolved radial profile under each $h$ assumption with the procedure described in Section~\ref{sec:fave-radial} to simulate a model image and evaluate $P(h)$. Section~\ref{sec:demonstration} will demonstrate that $P(h)$ is usually relatively smooth, so 10 to a few 10s of points in either linear or log-space for reasonable values of $h$ ($<$1) are generally sufficient to sample the distribution and interpolate between the points sampled. If the width of $P(h)$ is sufficiently narrow, we may place meaningful lower and/or upper constraints on $h$ using the 16th and 84th percentiles, and use the median to estimate the best-fit value of $h$. The best-fit $h$ estimate is then used as the height assumption to re-fit the radial profile, this time optimised for $h$. If the distribution of $P(h)$ is too wide to provide meaningful constraints on $h$ (e.g., $P(h)$ does not tend towards 0 at $h \gtrsim 0.3$), this would imply that the fitted model is either not sensitively dependent on height assumption due to the disk being very close to face-on, or that the noise is too large. In such a scenario, we proceed to fit final radial profile assuming a flat disk (i.e., $h = 0$) or any other reasonable value. 


\subsection{Extension to scattered light observations}
\label{sec:scattered_light}

\begin{figure}
    \centering
    \includegraphics[width=7cm]{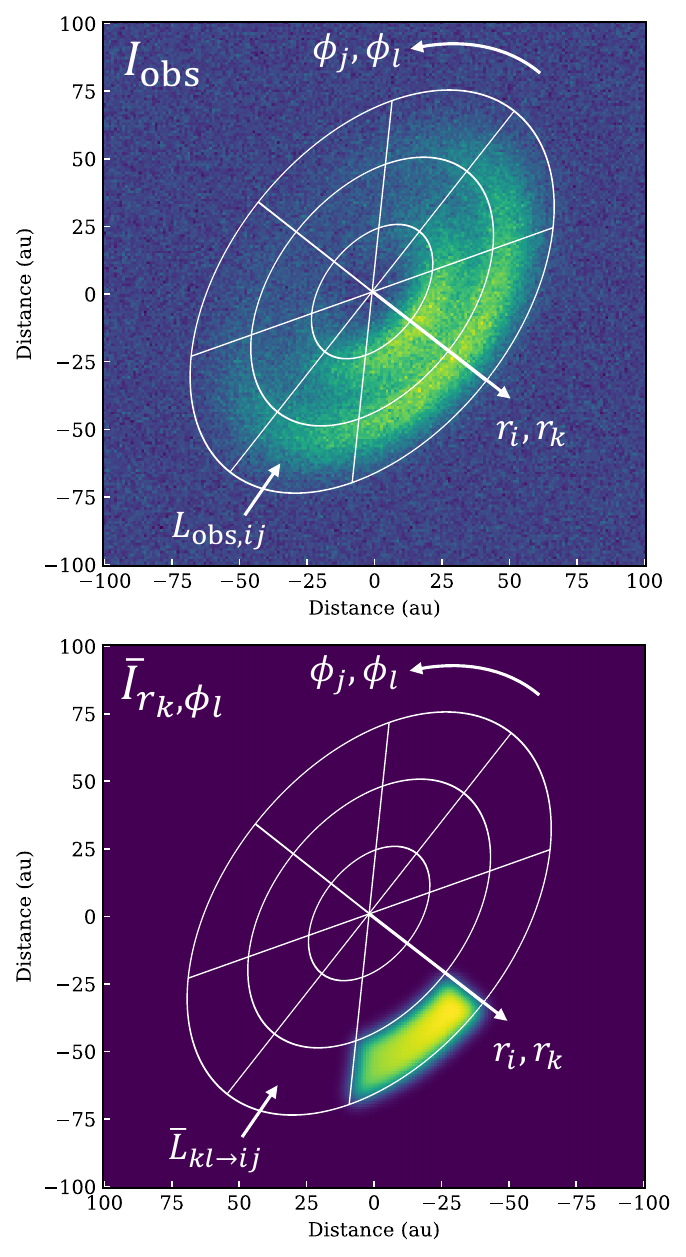}
    \caption{Schematic diagram showing the coordinate system applied to the observed image (top) and a model image of an $(r, \phi)$ element (bottom). }
    \label{fig:scfave}
\end{figure}

\subsubsection{Full theoretical treatment}
The construction of \texttt{rave} so far assumes that emission by particles in the disk is isotropic. This is satisfied for thermal emission from dust grains in the disk, which is typically probed at mid-infrared wavelengths and longer. However, at optical and near-infrared wavelengths, thermal emission is too weak to be detected. Instead, these shorter wavelengths primarily detect light from the host star scattered towards the line of sight by dust grains in the disk. However, unlike thermal emission which is isotropic, light that is scattered by dust particles is anisotropic, resulting in disk images that no longer appear to be azimuthally symmetric even if the underlying surface density distribution is. Given that these scattered light images of debris disks are also crucial to understanding their structure, ideally we would also be able to recover their radial profiles non-parametrically from scattered light observations. 

We can write the scattered light surface brightness (without PSF convolution) as
\begin{equation}
    \label{eq:sc}
    I_\text{sc}(r, \phi) = \frac{F_* d^2}{4 \pi r^2} \tau(r) \alpha_0 \eta(r, \theta),
\end{equation}
where $F_*$ is the stellar flux density, $d$ is the distance to the observer, $\tau$ is the optical depth of the dust disk, $\alpha_0$ is the albedo, $\eta$ is the scattering phase function and $\theta$ is the scattering angle. Note that we have defined the albedo and scattering phase function such that $\int_0^{\pi}d\theta\int_0^{2\pi}d\phi'' \eta \sin(\theta) = 4\pi$,
and that the azimuthal angle in the disk, $\phi$, is related to the scattering angle, $\theta$, by
\begin{equation}
    \label{eq:scattering_angle}
    \cos{\theta} = \cos{\phi} \sin{\iota},
\end{equation}
where $\phi=0$ corresponds to the direction of the line of ascending node and $\theta=0$ corresponds to forward scattering. 
The specific form of $\eta$ depends on factors such as the dust grain's composition and porosity, but generally scattering is more efficient in the forward direction, resulting in observed images that appear brighter in the half of the disk that is closer to us (i.e., in front of the line of nodes). 

In reality, the scattering phase function and albedo do not have to remain constant across all locations in the disk. It is in theory possible to simultaneously fit to the scattering phase function at every radius, $\eta(r, \theta)$, and the radial surface density profile multiplied by the albedo, $\alpha_0(r) \tau(r)$, given a disk image. To describe this complexity in our existing non-parametric annuli modelling framework, we discretise the scattering phase function, $\eta$, analogous to the setup in Section~\ref{sec:fave}, but here partitioning it into not only $N$ radial bins (indexed by $\iota$) but also $M$ azimuthal bins (indexed by $j$), effectively transforming an image into an $N \times M$ matrix, with each entry given by
\begin{equation}
    L_{ij} = \frac{2}{(\Phi_j - \Phi_{j-1})(R_{i}^2 - R_{i-1}^2)} \int_{\Phi_{j-1}}^{\Phi_{j}} \int_{R_{i-1}}^{R_i} I_\text{sc}(r, \phi) r dr d\phi,
    \label{eq:l_ij}
\end{equation}
where $\vb* R$ is the radial boundaries of each annulus as before and $\vb* \Phi = (0, \phi_1, \dots, \phi_{N-1}, 2\pi)$ is the azimuthal boundaries of each azimuthal element within each annulus. 
A diagram showing the transform of the image to discretised polar coordinates is shown in Fig.~\ref{fig:scfave}. 

Since each point in the disk contributes flux everywhere in the image due to convolution with the PSF, to disentangle the radial profile and scattering phase function at each radius, we must recover the de-projected and deconvolved surface brightness at each point in the disk. The transform defined by Eq.~\eqref{eq:l_ij} can be applied to both the observed image of the disk to obtain $L_{\text{obs}, ij}$ and the image of each individual normalised $(r_k, \phi_l)$ element, which represents an arc-like segment of a given normalised annulus with a surface brightness of unity (indexed by $k$) within a given azimuthal bin (indexed by $l$), to obtain $\bar L_{kl \rightarrow ij}$. Note that this notation uses $(i, j)$ to indicate the region in the image and $(k, l)$ to indicate the element of the disk, such that $\bar L_{kl \rightarrow ij}$ means the flux contribution from the $(r_k, \phi_l)$ element in a disk to the $(r_i, \phi_j)$ region if the element were to have a surface brightness of unity, where we have defined the elements of the disk to have the same $\vb* R$ and $\vb* \Phi$ boundaries as the regions in the image. 

Substituting $L$ for $I$ into Eq.~\eqref{eq:sc}, we can write
\begin{equation}
    L_{\text{obs}, ij} = \frac{F_*}{4 \pi} \sum_{k=1}^{N} \sum_{l=1}^{M} \frac{\tau^\prime_k}{r_k^2} \eta_{kl} \bar L_{kl \rightarrow ij},
    \label{eq:4d}
\end{equation}
where $\tau^\prime_k$ is the surface density multiplied by the albedo at $r_k$ and $\eta_{kl}$ the scattering phase function at $(r_k, \phi_l)$.
As the observed surface brightness across all regions of the disk provides $N \times M$ such equations, it is possible to constrain all $N \times M$ entries of the quantity $\alpha_0 \tau_k \eta_{kl}$, effectively achieving deconvolution and de-projection of the image in 2D. Applying the normalisation of $\eta$, we may then solve for $\eta_{kl}$ and $\tau_k$ to within a constant factor, thereby recovering the distribution of both the radial surface density profile and the scattering phase function across radius. 

While such an approach is workable in theory, we speculate that it is unlikely able to produce meaningful results due to practical considerations. As \citet{Han2022} have found for the case of solving only for the radial profile, noise in the image and causes unphysical structures such as oscillations to be recovered when applying the inverted 2-dimensional matrix in Eq.~\eqref{eq:matrix_inversion}, which is then mitigated by randomising the annuli boundaries through a Monte Carlo approach. In the case of solving for both the radial and azimuthal profiles, the 4-dimensional nature of the problem as implied by $\bar L_{kl \rightarrow ij}$ in Eq.~\eqref{eq:4d} could severely worsen the issue to the extent that even mitigating unphysical oscillations with the randomisation of the radial and azimuthal boundaries may be challenging. Given that in most scattered light images of debris disks, sufficient signal-to-noise ratio (S/N) is only reached within a relatively narrow range of radii, we find the full treatment of the problem by attempting to recover both the $r$ and $\phi$ dependence of the scattering phase function to be unnecessary. We therefore proceed to propose a simplified approach. Future studies may implement and test the full theoretical treatment described in this section should a significant volume of scattered light observations with very high S/N become available.

\subsubsection{Simplification}
To simplify the fitting required to recover the de-projected and deconvolved radial profile from scattered light images, we assume that the scattering phase function is independent of radius. We then perform the non-parametric fitting over two steps, first to derive an empirical scattering phase function, $\eta(\theta)$, directly from the observations, before proceeding to fit the radial profile following the procedure described in Section~\ref{sec:fave} except accounting for the $\eta(\theta)$ in each fitting annulus, $I_{\text{ann-}i}$. 

To derive the scattering phase function in step one, we adopt a similar approach to that described in \citet{Olofsson2020} by first deriving the radially-averaged azimuthal profile,
\begin{equation}
    K_\text{obs}^{(r)}(\phi) = \frac{2}{R_N^2}\int_{0}^{R_N} I_\text{obs}(r, \phi) r dr.
\end{equation}
For a smooth scattering phase function relative to the PSF size, the azimuthal profile offers a good empirical estimate of the scattering phase function for the purpose of retrieving the radial profile, as we will demonstrate in Section~\ref{sec:scdemo}. Assuming $\eta(\theta) \propto K_\text{obs}^{(r)}(\theta)$, we could then constrain $\eta(\theta)$ to within a constant factor which is degenerate with $\alpha_0\tau(r)$, since only a subset of scattering angles are present in the disk at $i < 90^\circ$ such that $\eta(\theta)$ cannot be normalised. Nonetheless, we may still proceed to recover the functional form of $\tau(r)$ despite the degeneracy of this constant factor. 

In step two, we apply \texttt{rave} while accounting for the scattering phase function estimated in step one, simulating each annulus used for fitting with this $\eta(\theta)$ and at the same inclination as the disk and at the assumed vertical aspect ratio $h$. The remainder of the procedure in Section~\ref{sec:fave} still applies, with the only difference being that $I_{\text{ann-}i}$ is simulated under the scattering phase function derived. 

Note that the azimuthally averaged profiles, $K^{(\phi)}$, as defined in Eqs.~\eqref{eq:kphi-obs} and \eqref{eq:kphi}, and thus $L^{(\phi)}$, do not need to be averaged over the entire azimuthal range. The primary reason for averaging over the entire azimuthal range is to improve S/N. As scattered light observations are known to exhibit significant artefacts due to the optics and PSF subtraction, diffraction spikes may still affect various azimuths of the disk image even when the central PSF artefacts are masked. In such circumstances, it may be more effective to average only a selected range of azimuths which do not appear to be heavily affected by artefacts in the image.

\subsection{Implementation}
The face-on extension to the algorithm with scattered light support is implemented in the objected-oriented \texttt{Python} package, \texttt{rave}. Installation instructions and demonstrations are available at \href{https://github.com/yinuohan/rave}{github.com/yinuohan/rave}.

\section{Demonstration}
\label{sec:demonstration}

\begin{figure*}
    \centering
    \includegraphics[width=11.2cm]{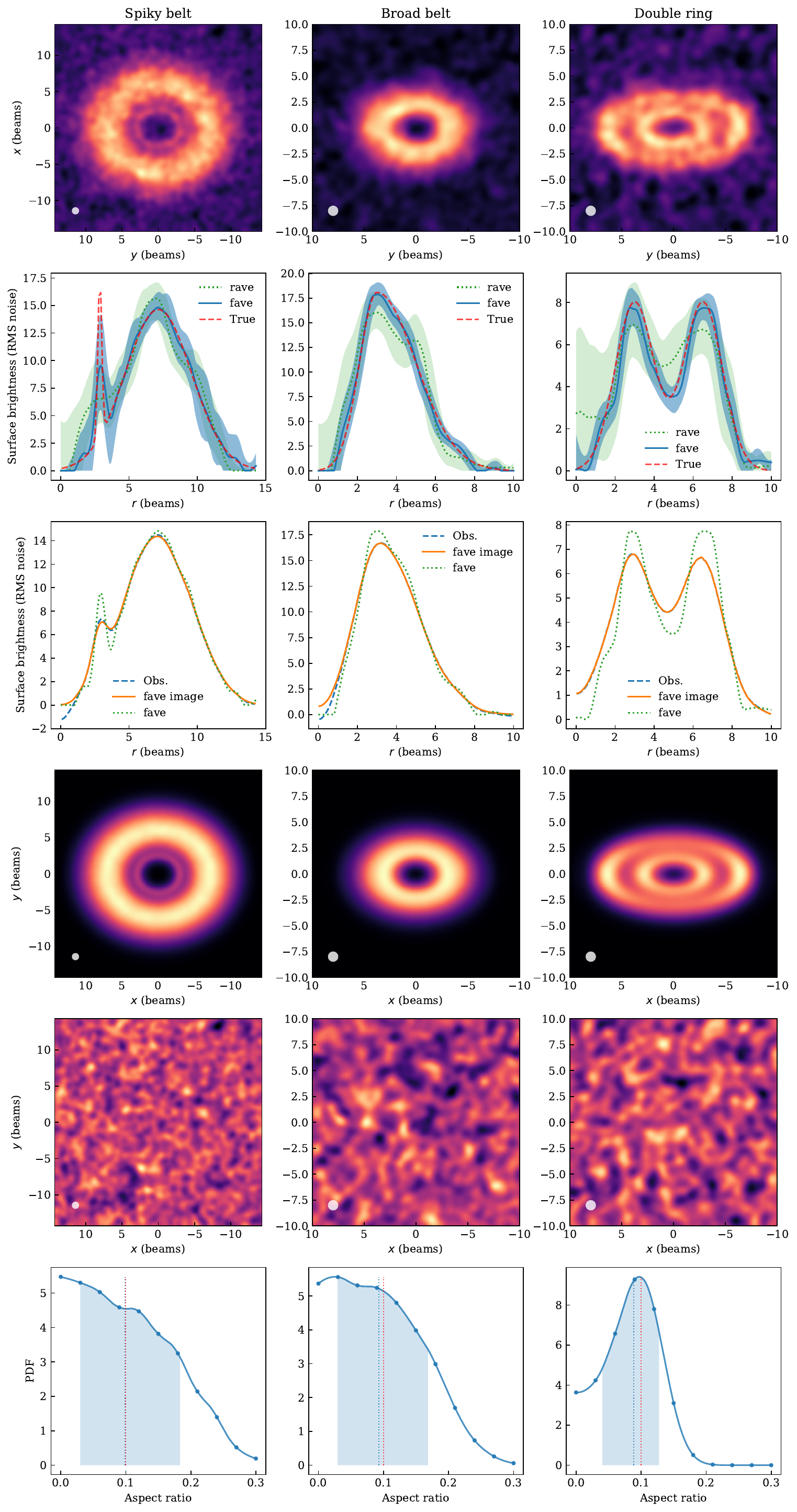}
    \caption{Demonstration of \texttt{fave} applied to simulated observations. Row 1: simulated images being fitted to. Row 2: radial profiles de-projected and deconvolved non-parametrically  with \texttt{rave} and \texttt{fave}. The true profile is overplotted. Row 3: The azimuthally averaged profile of the simulated observations (``Obs.''), beam-convolved model image (``fave image'') and de-projected and deconvolved radial profile (``fave''). Row 4: Beam-convolved image of the best-fit model. Row 5: residual map (simulated observations from row 1 $--$ model image from row 4). Row 6: probability density distribution of the vertical height aspect ratio inferred from the $\chi^2$ of the best-fit model assuming a range of aspect ratios. The true scale height is indicated with a red dotted line, and the median and 16th to 84th percentile interval with a blued dotted lines and shaded region. The scale height is not well-constrained for the test cases at 30$^\circ$ and 45$^\circ$ inclination, but useful constrains are derivable for the 60$^\circ$ inclination test case.}
    \label{fig:demos}
\end{figure*}

\begin{figure*}
    \centering
    \includegraphics[width=18cm]{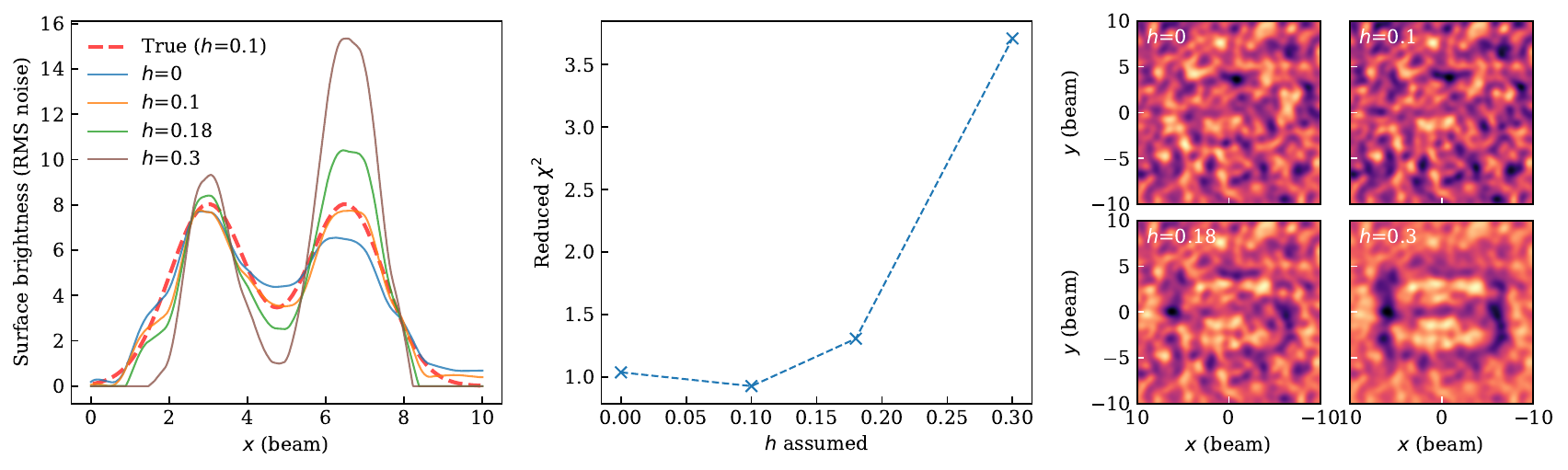}
    \caption{Demonstration of the mechanisms of scale height inference. Left: Radial profiles recovered under a range of different height assumptions. Middle: Reduced $\chi^2$ of the residual images as a function of aspect ratio assumed when fitting the radial profile. Right: residual images of the models fitted under the different height assumptions. The model with the lowest residuals is the one fitted under the true height assumption, thereby offering a way to constrain the scale height aspect ratio. This model also most accurately reproduces the true radial profile. }
    \label{fig:demos_height}
\end{figure*}


This section describes the performance of the extended algorithm by fitting to simulated observations generated with a known surface brightness profile and vertical height. The test cases involved include a ``spiky disk'', ''broad disk'' and ''double ring'', whose surface brightness profiles are defined by

\begin{equation}
    \label{eq:testcase1}
    f_\text{spiky}(r) = g(r, 7.1, 2.4) + g(r, 2.9, 0.1),
\end{equation}
\begin{equation}
    \label{eq:testcase2}
    f_\text{broad}(r) = \begin{cases}
    g(r, 3, 0.9), \ 0 \leq r < 3,\\
    g(r, 3, 2),   \ r \geq 3,
    \end{cases}
\end{equation}
\begin{equation}
    \label{eq:testcase3}
    f_\text{double}(r) = g(r, 3, 1) + g(r, 6.5, 1),
\end{equation}
\noindent where
\begin{equation}
    g(r, \mu, \sigma) = \exp(-\frac{(r-\mu)^2}{2\sigma^2}).
\end{equation}
The unit of $r$ is the full width at half maximum (FWHM) of the beam/PSF.

In this section, we simulate the three test cases both for ALMA-like observations and scattered light-like observations to evaluate the performance of the algorithm to recover the radial profiles in both contexts as well as to constrain the scale height. For clarity, we refer to the extended algorithm described in Section~\ref{sec:fave} as \texttt{fave} (replacing the ``r'' with ``f'', as it is optimised for face-on disks) and the original edge-on version of the algorithm as \texttt{rave}.

\subsection{Thermal emission}
\label{sec:almademo}

We simulated thermal emission images of the three test cases in the style of ALMA observations (imaged with the \texttt{clean} algorithm, \citealp{Hogbom1974}) by introducing correlated noise to the images. The images were simulated with a mean signal-to-noise ratio (S/N) per beam of 5 across the region with disk emission, which was achieved with a RMS noise per beam of 0.07, 0.06 and 0.12 respectively for the spiky disk, broad disk and double ring test cases in units of the radial profile defined in Eqs.~\eqref{eq:testcase1} to \eqref{eq:testcase3}. The inclination was set to be 30$^\circ$, 45$^\circ$ and 60$^\circ$ degrees respectively and the scale height aspect ratio, $h = H(r)/r$, was set to be constant and equal to 0.1 for all test cases. The simulated images and the underlying radial profiles for the three test cases are shown in Fig.~\ref{fig:demos} rows~1 and 2.

\subsubsection{Radial profile}
\label{sec:almademoradial}

As the de-projection and deconvolution approach developed in this study assumes axisymmetry, the first step before applying the algorithm is to check whether the disk image is indeed symmetric at the sensitivity and resolution of the observations. At millimeter wavelengths, systems which show unambiguous asymmetries in the dust continuum appear to be uncommon at the sensitivity and resolution of current observations, which could be checked by, for instance, self-subtracting the disk image about the star and about the major axis of the disk.

Upon confirmation of symmetry, we fitted the radial surface brightness profiles of each of the three test cases using both \texttt{rave} and \texttt{fave}. The \texttt{rave} profiles were fitted with $N$=10, 10 and 7 annuli for the spiky disk, broad disk and double ring test cases respectively, whereas the \texttt{fave} profiles were fitted with $N=30$, 15 and 15. The value of $N$ was chosen based on the principles described in Section~3.2.5 in \citet{Han2022}, which corresponds to the largest $N$ (and therefore highest spatial resolution) that is still robust against oscillatory artefacts due to resolution and noise level limitations of the image. While the \texttt{rave} profiles are independent of height assumptions, the \texttt{fave} profiles were fitted assuming the true aspect ratio, $h=0.1$. We defer the discussion on obtaining height constraints with \texttt{fave} in the absence of knowledge on the scale height to Section~\ref{sec:almademoheight}. All profiles were fitted with 100 iterations of randomised annuli boundaries and the resulting median profiles were smoothed with a Savitzky--Golay filter \citep{Savitzky1964}, which constitute the ``fitted profiles''. 

The fitted profiles are displayed in Fig.~\ref{fig:demos} row~2. We observe that the \texttt{fave} profiles exhibit excellent agreement with the true profile, demonstrating that the extended algorithm is able to successfully deconvolve and de-project the disk images even in the presence of noise. All major features in each test case are detected, including the two rings separated by a gap in the double ring test case, the asymmetric inner and outer edge in the broad belt test case and the Gaussian ring with a sharp perturbation on the inner edge in the spiky belt test case. 

The shaded regions in Fig.~\ref{fig:demos} row~2 indicate the uncertainties, which were obtained using the 16th and 84th percentiles of the profiles from the 100 interactions of individual fits with randomised annuli boundaries. Unlike the uncertainties in parametric models which are conditioned upon the assumed functional form of the radial profile being true, the uncertainties here are independent of such assumptions and reflect the band of possible models within which the true profile likely exists. As such, the shape of the true radial profile may deviate from the median profile shown in solid lines, such as by containing sharper features made ambiguous by noise and convolution with the beam/PSF. For example, the sharp perturbation in the spiky belt test case is difficult to reproduce perfectly as its width is only one third of the beam size, but this feature was still fitted to be at the correct radial location and the increased uncertainties in this region largely contain the true spike. 

The \texttt{rave} profiles were also able to recover the structures, albeit with larger uncertainties. The increased precision of \texttt{fave} relative to \texttt{rave} in non-edge-on systems is unsurprising given that \texttt{rave} fits to the vertically summed profile, $K_\text{obs}^{(y)}(x)$, which much more easily drowns out small-amplitude features with noise, such as for the narrow inner ring in the spiky belt test case, whereas such features are better preserved in the azimuthally averaged profile of the 2D disk emission, $K_\text{obs}^{(\phi)}(r)$, which \texttt{fave} takes advantage of. 

In fact, although \texttt{rave} fits are independent of inclination and should deliver constant performance without noise, such a more face-on configuration, in the presence of noise, should result in worse performance for \texttt{rave} since more noise is present across a face-on disk image compared to an edge-on perspective. On the other hand, \texttt{fave} is unsuitable for edge-on disks which are radially unresolved along the minor axis, but performs better when the disk is closer to face-on. The strengths of the two methods are therefore complimentary, and we suggest that radially unresolved disks along the minor axis are better suited for fitting with \texttt{rave}, whereas \texttt{fave} is expected to deliver better performance otherwise.

Fig.~\ref{fig:demos} row~3 displays the azimuthally averaged profiles,  $K_\text{obs}^{(\phi)}(r)$, which is the observable that \texttt{fave} fits to. The azimuthally averaged profiles of the fitted models, which are viewed at the appropriate incination and convolved with the beam, provided a highly consistent fit to the simulated observations. To visualise the deconvolvution and de-projection capapabilities of \texttt{fave}, the fitted radial profiles are also plotted in dotted lines, which contain sharper structures which are smoothed out by the beam to give the observable profiles. This is especially pronounced for the spiky belt, in which case the spike only manifests as a small bump in the observable but requires a significantly higher degree of ``spiky-ness'' to reproduce. The fact that the observable can be accurately reproduced with an underlying spike that is still less sharp than the true profile reflects that very sharp features that are much smaller than the beam size are difficult to accurately constrain from images as expected. 

Finally, the 2D beam-convolved model images are shown in Fig.~\ref{fig:demos} row~4, and the 2D residuals in row~5 indicate a highly consistent fit to the simulated observations, with no significant residual noise structures. These test cases together demonstrate that \texttt{fave} is able to successfully model the radial structure of optically thin disks, offering improved deconvolution and de-projection performances over the original \texttt{rave} approach for non-edge-on disks. 

In general, applying \text{fave} to interferometric observations, such as ALMA observations, requires first imaging the visibilities. Imaging algorithms such as \texttt{clean} require a choice of imaging parameters, most notably the robust value that affects the balance between resolution and noise. Testing \texttt{fave} on clean images of the same dataset but imaged with different robust values shows no significant difference in the recovered profile for reasonable robust values (e.g., 0.5 or 2.0), since \texttt{fave} bins the disk image into discrete annuli that are generally comparable to or larger than the beam size. Despite the lack of sensitive dependence of the de-projection and deconvolution on imaging parameters, it could be beneficial to carry out the fit to different images of the same dataset if the observations are of relatively poor quality and the trade-off between resolution and sensitivity particularly constraining.

\subsubsection{Scale height}
\label{sec:almademoheight}
The radial profiles in Section~\ref{sec:almademoradial} were fitted with \texttt{fave} assuming the true scale height for each image. However, the scale height may not be known \textit{a priori} when fitting to real observations. Depending on the resolution, belt width and S/N, the scale height aspect ratio could be constrained in a disk even if not edge-on \citep{Marino2016}. In practice, the scale height of most debris disks are found to be small \citep{Terrill2023} and the effect of scale height assumptions on the recovered radial profile is expected to be negligible compared to other sources of uncertainties, such as noise, such that assuming a flat disk ($h$=0) is sufficient to provide accurate recovery of the radial profile. However, the scale height is a key property characterising the structure of debris disks, and some disks do have scale heights that appear to be larger than 0.1 \citep{Matra2019, Terrill2023}, in which case scale height assumptions could matter when fitting to the radial structure. 

An example showing the scale height dependence of the fitted radial profile is shown in the left panel of Fig.~\ref{fig:demos_height}. As the assumed scale height increases and spreads out the flux vertically along the minor axis, the radial profile becomes sharper to compensate for this effect along the radial direction, particularly at larger radii where the scale height is larger under a given aspect ratio. 

In such scenarios where the radial profile significantly co-varies with height, it is possible to leverage this behaviour and place constraints on the scale height of non-edge-on disks when applying \texttt{fave}. While the effects of scale height on the azimuthally averaged profile can be partly compensated for by changes in the radial profile, the radial and vertical effects are rather different in 2D, where the disk image can only be fully reproduced when the scale height assumption is correct. This is demonstrated in the right panels of Fig.~\ref{fig:demos_height}, which show the residual images of the four models in the left panel fitted under different height assumptions. The model fitted under the true scale height ($h = 0.1$) produces a good match to the observed 2D image, whereas significantly different scale height assumptions result in fits with significant residual structures. The reduced $\chi^2$ of these models as a function of the scale height assumed are shown in the middle panel of Fig.~\ref{fig:demos_height}, which indeed reaches a minimum when the scale height assumption is correct. The radial profile recovered under this correct assumption is also the most accurate, as shown in the left panel. 

To evaluate the height constraints that can be derived from the three test cases, we fitted the radial profile of each test case with \texttt{fave} under 11 different height assumptions evenly spaced between 0 and 0.3. We simulated the resulting model image under each test case to compute the $\chi^2$ for each model. Assuming that the squared residuals follow a $\chi^2$ distribution, we may convert the $\chi^2$ values into a probability density distribution, with $P(h) \propto \exp[\chi^2(h) / 2]$. The resulting probability density distribution for the aspect ratio in each test case is shown in Fig.~\ref{fig:demos} row~6. 

We find that the scale height reaches a clear peak for the double ring test case at $h = 0.097$, where the true value is 0.1, with the median, 16th and 84th percentiles estimating $0.09_{-0.05}^{+0.04}$. Indeed, this is the most edge-on test case of the 3, at an inclination of 60$^\circ$, for which we expect to obtain better constraints on the height that the other examples. The height constraints are expected to be further narrowed for even more inclined disks. For the other two test cases, which are at inclinations of 30$^\circ$ and 45$^\circ$, the vertical distribution is not as readily accessible, and it is only possible to rule out a scale height of beyond $\sim$0.25. 
For these two examples, we find that the radial profiles fitted for height assumptions that are smaller than $\sim$0.15 are also highly consistent with each other. The lack of sensitive dependence on the scale height for these test cases is expected from the more face-on viewing geometry, under which the image is less affected by the vertical distribution, and this is also reflected in the relatively constant $\chi^2$ values over a broad range of relatively low scale heights. 

In summary, constraints on the vertical height may be obtained by fitting to the radial profile with \texttt{fave} over a range of height assumptions and comparing the resulting 2D models to the observations. For more edge-on disks, this offers an estimate of the vertical height to inform the more likely scale height, the radial profile fitted assuming which should be interpreted as the more likely radial profile. For more face-on disks, the radial profile could be less sensitive to the height assumption, in which case any scale height within a reasonable range characterised by relatively constant $\chi^2$ could serve as an appropriate height assumption under which to derive the radial profile.

\subsection{Scattered light}
\label{sec:scdemo}

\begin{figure*}
    \centering
    \includegraphics[width=11.2cm]{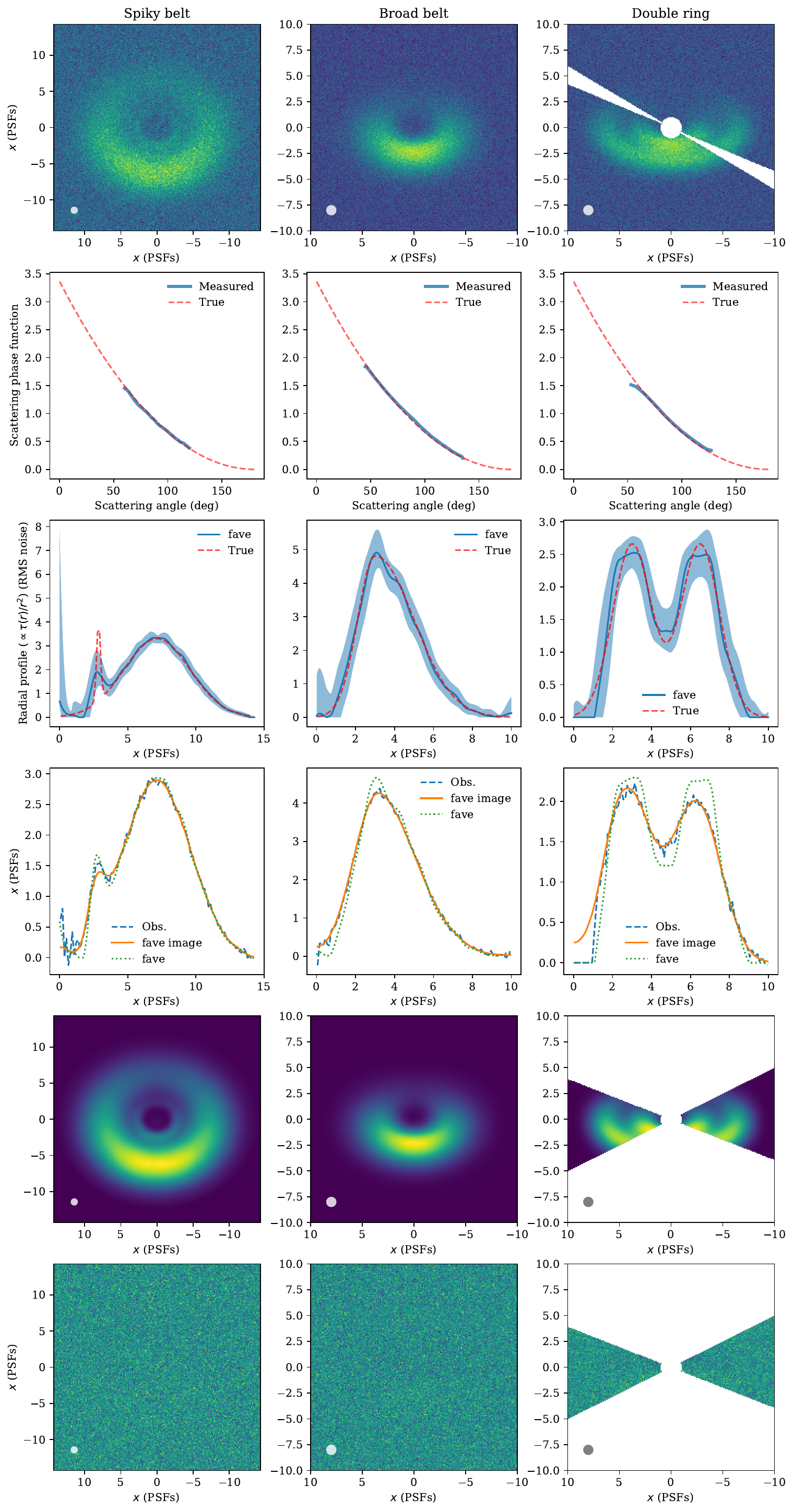}
    \caption{Demonstration of the scattered light extension to \texttt{fave} applied to simulated observations. Row 1: simulated images being fitted to. Parts of the double ring test case are masked to the information lost due to PSF subtract artefacts and diffraction spikes. Row 2: true scattering phase function and that emperically derived from the observations. Row 3: radial profiles de-projected and deconvolved non-parametrically  with \texttt{rave} and \texttt{fave}. The true profile is overplotted. Row 4: The azimuthally averaged profile of the simulated observations (``Obs.''), beam-convolved model image (``fave image'') and de-projected and deconvolved radial profile (``fave''). Row 5: Beam-convolved image of the best-fit model. Note that a smaller range of azimuths than is shown outside the mask in row 1 was used for fitting the double ring test case due to the central circular mask obstructing part of the inner disk emission, hence the larger masked out region in the model image. Row 6: residual map (simulated observations from row 1 $--$ model image from row 5).}
    \label{fig:demos_spf}
\end{figure*}

To evaluate the performance of \texttt{fave} when accounting for the effects of the scattering phase function, we simulated scattered light images of the three test cases assuming a scattering phase function given by
\begin{equation}
    \label{eq:spf}
    \eta(\theta) = \frac{2}{\pi^2 - 4} (\pi - \theta)^2.
\end{equation}
We simulated images of the test cases in Eqs.~\eqref{eq:testcase1} to \eqref{eq:testcase3} defined in spatial units of PSFs by projecting them onto identical 200 by 200 pixel grids and adding independent noise per simulated detector pixel. This corresponds to PSF FWHM over-sampling factors of 7, 10 and 10 for the spiky disk, broad disk and double ring test cases respectively. The images were simulated with a mean S/N per pixel of 1 across the region with disk emission, which was achieved with an RMS noise per pixel of 0.38, 0.21 and 0.30 for the three test cases respectively in units of the radial profile defined in Eqs.~\eqref{eq:testcase1} to \eqref{eq:testcase3}. The inclinations of the three test cases are 30$^\circ$, 45$^\circ$ and 60$^\circ$ degrees as before and the scale height aspect ratios are 0, 0.05 and 0.1. 

To test the performance of the algorithm in the presence of instrumental or PSF subtraction artefacts that extend into the region of disk emission at certain azimuths, we masked out regions in the simulated image of the most inclined test case, the double ring. The regions that are masked out include a central circular region with twice the diameter of the PSF FWHM, which is analogous to the influence of PSF subtraction artefacts that omit information near the PSF core, and the range of azimuths within the intervals of (130$^\circ$, 140$^\circ$) and (310$^\circ$, 320$^\circ$) in the disk plane, with 0$^\circ$ being the ascending node (to the right of the image), which simulates information lost from diffraction spikes or instrumental artefacts. The simulated images are shown in Fig.~\ref{fig:demos_spf} row~1. Here, we have not simulated the effects of PSF over- or under-subtraction, which could be mitigated by observing with polarisation differential imaging or using PSF subtraction techniques such as non-negative matrix factorisation \citep{Ren2018}. 

The first step in fitting to scattered light images with \texttt{fave} is to empirically recover the scattering phase function, $\eta(\theta)$. For each image, we approximated $\eta(\theta)$ with the radially averaged azimuthal profile of the disk, $K^{(r)}_\text{obs}(\phi)$, where the scattering angle, $\theta$, and the azimuthal angle in the disk plane, $\phi$, are linked by Eq.~\eqref{eq:scattering_angle}. 

Since none of the disks are perfectly edge-on, it is not possible to sample the full range of scattering angles from 0 to $\pi$, which prevents us from deriving the appropriate normalisation factor for $\eta(\theta)$. It is therefore only possible in practice to obtain the scattering phase function scaled by a constant factor, $k$. In the context of the test cases, since the true scattering phase functions are known, we have scaled the smoothed azimuthal profiles, $K^{(r)}_\text{obs}(\theta)$, by the appropriate factor, $k$, for ease of comparison. 

The true scattering phase function and those measured from the simulated observations are displayed in Fig.~\ref{fig:demos_spf} row~2. Across the three test cases, $K^{(r)}_\text{obs}(\theta)$ offers an accurate proxy for $\eta(\theta)$ within the range of scattering angles sampled within the disk image. For the double ring image, the masked out PSF region extends into the inner ring along the minor axis, so including these azimuths could bias subsequent radial profile recovery. We therefore only consider the regions of the image in the azimuthal intervals of (-38$^\circ$, 45$^\circ$) and (142$^\circ$, 225$^\circ$) in the disk plane, which avoids the masked out diffraction spikes while being primarily centred along the major axis of the disk where the effect of the masked out PSF core on the disk emission is lower. The limited azimuthal range reduces the range of scattering angles for which the scattering phase function can be measured. However, the masked out regions should still in theory contribute flux to the unmaksed regions due to the effect of the PSF. In subsequent fitting, we assumed the scattering phase function to be constant and equal to the boundary values within the masked regions when generating model annuli to fit the radial profile. 

Using the scattering phase functions derived, we fitted the radial profiles of the test cases with \texttt{fave}, using $N=20$, 15 and 10 for the test cases respectively. The results are displayed in Fig.~\ref{fig:demos_spf} row~3, which show that the radial structures have been correctly recovered, demonstrating that the algorithm can successfully de-project and deconvolve these idealised simulated scattered light images by taking into account the empirically derived scattering phase function. Note that since only $k\eta(\theta)$ is constrained for the scattering phase function, Eq.~\eqref{eq:sc} implies that the recovered radial profile corresponds to $(1/k) (F_* d^2 / 4 \pi) \alpha_0 \tau(r) / r^2$. In other words, the radial surface density profile, $\tau(r)$, can only be recovered to within a constant factor, which is due to both the unknown normalisation factor, $k$, for the scattering phase function, and the unknown albedo, $\alpha_0$, which is degenerate with the surface density in scattered light. In these test cases, we have applied the normalisation correction with knowledge of the true scattering phase function, as we did for the scattering phase function derived, and the true and fitted radial profiles plotted in Fig.~\ref{fig:demos_spf} row~3 both correspond to the quantity $(F_* d^2 / 4 \pi) \alpha_0 \tau(r) / r^2$.

Fig.~\ref{fig:demos_spf} row~4 shows the azimuthally averaged profile for the three test cases. Rows~5 and 6 of the figure show the simulated model images (convolved with the PSF) and the residual image. These diagnostics indicate that the model is able to accurately reproduce the simulated observations within noise, demonstrating that the fitting has proceeded successfully. For the double ring, the azimuthally averaged profile is only averaged over the narrow azimuthal intervals used for fitting described earlier in this section. It is interesting to note that while no information was given to the algorithm about the masked out regions as indicated in the model image and residuals, the fact that an appropriate model can still be derived for the unmasked region reflects that the relatively small PSF causes the flux contribution to remain relatively local. We also note that at the noise level for these simulated observations, it is becoming difficult to fully reproduce the spike in the azimuthally averaged profile of the spiky disk, but the lack of significant residual structures in the residual image indicate that the fitted model is still appropriate to explain the data.

\section{Application}
\label{sec:application}
In this section, we apply our non-parametric fitting method to real observations of a sample of debris disks to constrain their radial and vertical structures. We then use the results of this uniform analysis to search for any correlations that could reflect general properties of debris disks.

\subsection{Sample}
\label{sec:sample}

\begin{table*}
    \centering
    \caption{Summary of targets fitted with \texttt{rave} or \texttt{fave}.}
    \label{tab:targets}
    \begin{tabular}{lllllllllllll}
\hline\hline
Star        & Alias        & $d$ (pc) & $\iota$ ($^\circ$) & SpT  & $T_*$ (K) & Age (Myr)        & Method        & $N_r$ & $N_H$ & $y_\text{mid}$ (au) & Band & Reference \\ \hline
HD\,197481     &              & 9.7          & 88.4               & M1V  & 3597      & 22$\pm$3             & \texttt{rave} & 10    & 3     &  1.3                    & 6    & \citealp{Daley2019}          \\
HD\,15115     &              & 49           & 88                 & F4IV & 6740      & 23$\pm$3             & \texttt{rave} & 10    & 3     &  11                     & 6    & \citealp{Macgregor2019}          \\
HD\,110058    &              & 130          & 87                 & A0V  & 7710      & 17                   & \texttt{rave} & 7     & 3     &  17                     & 7    & \citealp{Hales2022}          \\
HD\,32297     &              & 132.8        & 87                 & A0V  & 7430      & 30                   & \texttt{rave} & 7     & 3     &  25                     & 6    & \citealp{MacGregor2018}          \\
HD\,61005     &              & 36.5         & 85.7               & G8V  & 5560      & 40                   & \texttt{rave} & 12    & 5     &  3.7                    & 6    & \citealp{MacGregor2018}          \\
HD\,50571     &              & 33.9         & 85                 & F5V  & 6560      & 300$\pm$120          & \texttt{rave} & 4     & 2     &  7.1                    & 7    & \citealp{Matra2025}          \\
HD\,35841     &              & 103.7        & 84                 & F3V  & 6340      & 40                   & \texttt{rave} & 5     & 2     &  5.3                    & 6    & \citealp{Matra2025}          \\
HD\,14055     & $\gamma$~Tri & 34.4         & 81.1               & A1V  & 9210      & 230$\pm$70           & \texttt{rave} & 7     & 3     &  35                     & 6    & \citealp{Matra2025}          \\
HD\,158352    &              & 63.8         & 81                 & A8V  & 7450      & 750$\pm$150          & \texttt{rave} & 5     & 2     &  26                     & 6    & \citealp{Matra2025}          \\
HD\,9672      & 49~Cet       & 57.1         & 79.1               & A1V  & 8750      & 40                   & \texttt{fave} & 10    &       &                         & 8    & \citealp{Higuchi2019}          \\
HD\,10647     & $q^1$~Eri    & 17.3         & 77.2               & F9V  & 6150      & 1600                 & \texttt{fave} & 10    &       &                         & 7    & \citealp{Lovell2021}          \\
HD\,109573    & HR~4796      & 70.8         & 76.5               & A0V  & 9980      & 8                    & \texttt{fave} & 15    &       &                         & 7    & \citealp{Kennedy2018}          \\
HD\,76582     &              & 48.8         & 72                 & F0IV & 7750      & $540_{-200}^{+1500}$ & \texttt{fave} & 10    &       &                         & 6    & \citealp{Matra2025}          \\
HD\,161868    & $\gamma$~Oph & 31.5         & 68                 & A1V  & 9070      & $185_{-135}^{+90}$   & \texttt{fave} & 10    &       &                         & 6    & \citealp{Matra2025}          \\
HD\,92945     &              & 21.5         & 65.4               & K1V  & 5194      & $170_{-50}^{+80}$    & \texttt{fave} & 20    &       &                         & 7    & \citealp{Marino2019}          \\
GJ\,14        &              & 14.7         & 64                 & K7   & 4100      & ?                    & \texttt{fave} & 10    &       &                         & 6    & \citealp{Matra2025}          \\
HD\,191089    &              & 50.1         & 60                 & F5V  & 6460      & 22$\pm$6             & \texttt{fave} & 7     &       &                         & 6    & \citealp{Matra2025}          \\
HD\,107146 &              & 27.47        & 20                 & G2V  & 5890      & $150_{-30}^{+50}$    & \texttt{fave} & 20    &       &                         & 7    & \citealp{Marino2018}          \\ \hline
\end{tabular}
\end{table*}

We modelled available ALMA observations of 18 debris disks, primarily based on those within a sample studied by \citet{Terrill2023}. This sample is ideal for studying the radial and vertical structure simultaneously as it contains the best resolved axisymmetric disks at high inclination. These disks have also been systematically modelled by \citet{Terrill2023} using the visibility-space de-projection and deconvolution algorithm, \texttt{frank}, allowing for cross validation and comparison between the results of two different approaches. Additionally, we also include resolved observations of the highly inclined disk, HD\,76582 \citep{Matra2025}, as well as high-resolution observations of HD\,107146 (Band~7, \citealp{Marino2018}) known for its complex multi-ring structure. The sub-millimeter galaxies in HD\,107146 and HD\,76582 have been subtracted to avoid biasing the fitted disk structure. Table~\ref{tab:targets} provides a summary of these targets.

For each disk, we used identical data reductions to those presented in \citet{Terrill2023} and assumed a position angle, inclination and stellar flux equal to those in Table~2 of the same paper. These values were determined by fitting parametric models with a Gaussian radial and vertical profile \citep{Matra2025}, except for cases where the radial profile deviates significantly from being Gaussian (HD\,92945, \citealp{Marino2021}) or where such an approach does not provide robust constraints (HD\,110058, \citealp{Esposito2020, Hales2022, Stasevic2023}), in which case values fitted from the corresponding references are used. For HD\,110058, scattered light and ALMA gas observations provide a range of different inclination estimates between 80$^\circ$ and 90$^\circ$. We adopt a value of 87$^\circ$ here, which lies within the credible range inferred from the GPIES scattered light survey \citep{Esposito2020} and ALMA CO observation \citep{Hales2022}. 

\begin{figure*}
    \centering
    \includegraphics[width=18cm]{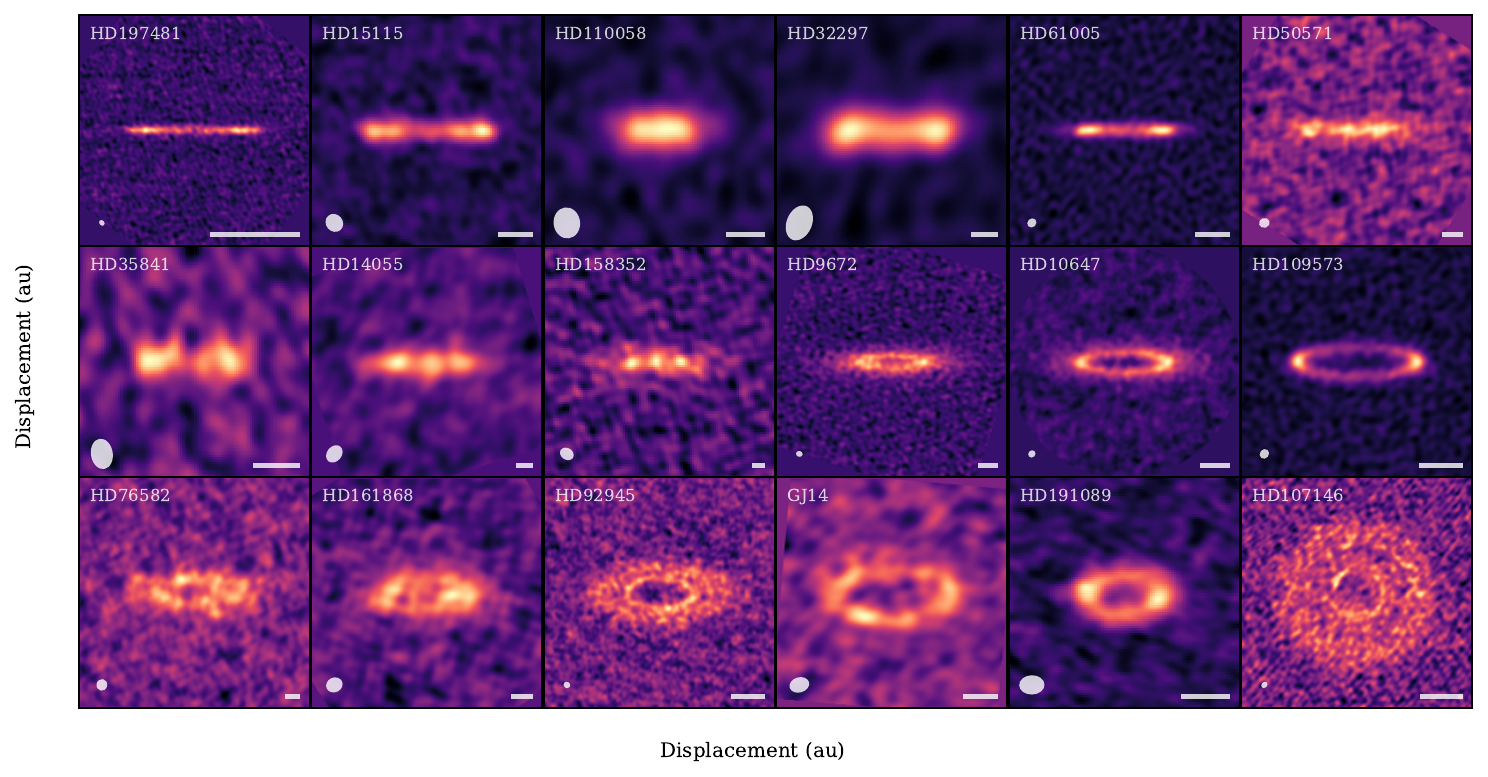}
    \caption{ALMA images of the sample of debris disks used in this study. The images are sorted by inclination and are rotated such that the major axis is horizontal. The scale bar in each panel indicates 50~au. References and observing bands are shown in Table~\ref{tab:targets}.}
    \label{fig:image}
\end{figure*}

Some disk images showed significant residual structures when subtracting the image rotated by 180 degrees about the centre from the original image. Assuming that the disks are axisymmetric and that these residual structures are caused by centering offsets, we re-centred each disk image by performing a grid search over the positional offset along the major and minor axes that minimises the squared residuals of the rotationally self-subtracted image. Typical offsets were small and on the order of several 0.01$^{\prime\prime}$, consistent with systematic pointing uncertainties and astrometric uncertainties. The centred, rotated and star-subtracted disk images are shown in Fig.~\ref{fig:image}. 

\subsection{Modelling}
\label{sec:modelling}
We divided the sample into two groups to model their radial profiles non-parametrically with either \texttt{rave} or \texttt{fave}. In Fig.~\ref{fig:image}, in which the disks are ranked by inclination, all disks after HD\,9672 (inclusive) have resolved minor axes, allowing for the azimuthally averaged profile to be derived and making them more suitable for modelling with \texttt{fave}. All disks before HD\,158352 (inclusive) are closer to edge-on and do not have their central cavities resolved, so we modelled their radial profiles with \texttt{rave} instead. 

\begin{figure*}
    \centering
    \includegraphics[width=18cm]{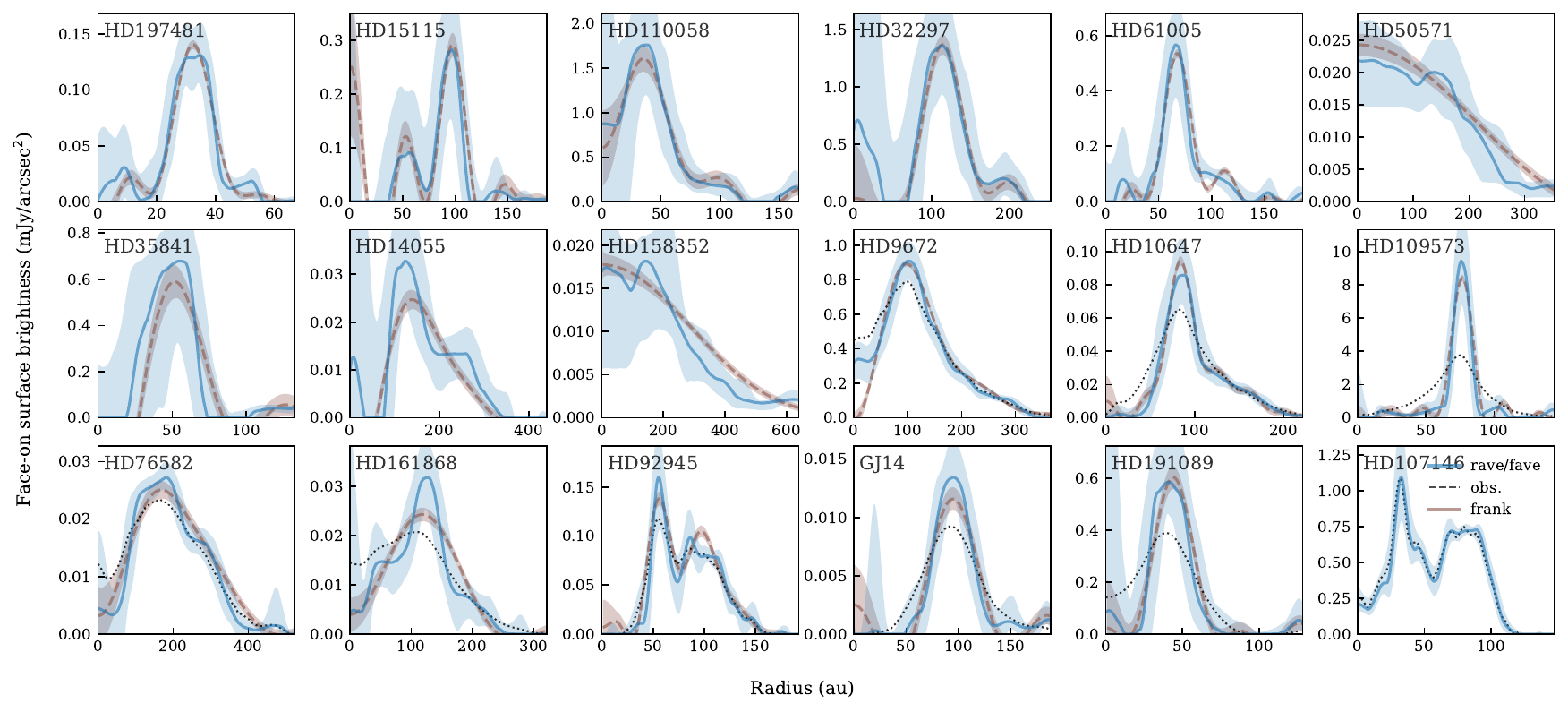}
    \caption{De-projected and deconvolved radial surface brightness profiles shown in solid blue lines, with the blue shaded regions showing the range of possible models. Disks before and including HD\,158352 which are closer to edge-on were fitted with \texttt{rave}, whereas all other disks were fitted with \texttt{fave}. For comparison, a version of the radial profiles fitted with \texttt{frank} in \citet{Terrill2023} provided by the authors are shown in brown dashed lines along with the 1$\sigma$ uncertainties. The azimuthally averaged profiles derived directly from the images are plotted in black dotted lines for the \texttt{fave} group, whereas such a profile cannot be meaningfully derived for the edge-on disks in the \texttt{rave} group. }
    \label{fig:rp}
\end{figure*}

The fitted radial profiles are presented in Fig.~\ref{fig:rp} and the number of annuli used for each disk, $N_r$, is shown in Table~\ref{tab:targets}. While the \texttt{rave} profiles are independent of height assumptions, the \texttt{fave} profiles displayed are those assuming the best-fitting aspect ratio. 

The aspect ratios were estimated using the same approach as demonstrated in Section~\ref{sec:almademoheight}. Specifically, we fitted \texttt{fave} profiles under 10 different aspect ratio assumptions, 0, 0.01, 0.02, 0.03, 0.04, 0.07, 0.12, 0.19, 0.31 and 0.5, which are logarithmically spaced with 0 included, simulating a model image under each to compute the $\chi^2$ value given the background noise level measured from the image. We then interpolated quadratically between these sample points and converted the $\chi^2$ values into a probability density distribution normalised to a probability mass of 1 assuming that the square residuals follow a $\chi^2$ distribution. Although the \texttt{rave} radial profiles are independent of height, we also simulated model images (which do depend on height) for disks in the \texttt{rave} group under different aspect ratio assumptions to constrain its value. 

\begin{figure*}
    \centering
    \includegraphics[width=18cm]{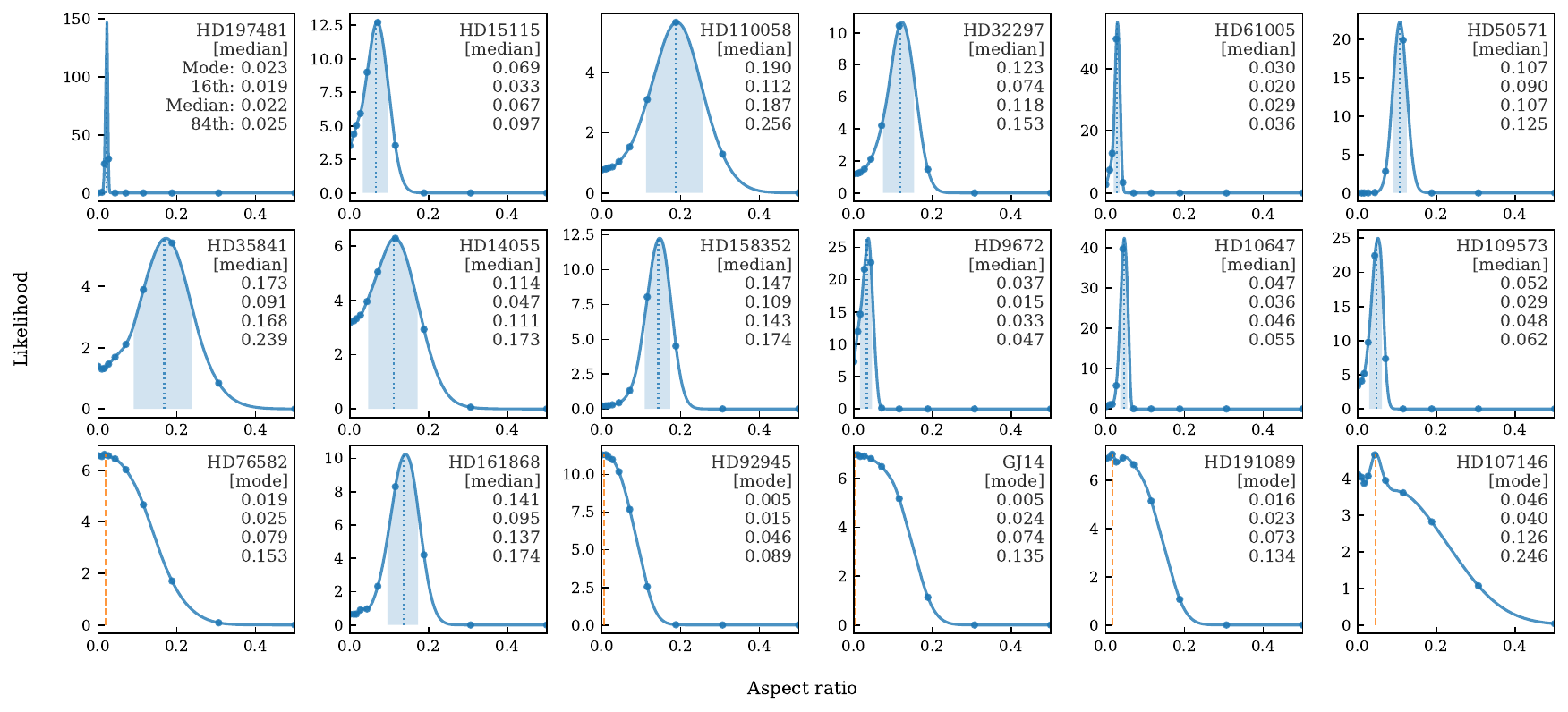}
    \caption{Probability density distributions of the aspect ratio of each disk inferred from model residuals assuming a range of aspect ratios. The text in each panel indicates the mode (maximum likelihood estimate), median and 16th and 84th percentiles of the distribution. Disks labelled with ``median'' have both upper and lower uncertainties on the aspect ratio, as indicated with shaded regions, and the dotted lines indicate the median of the distributions. Disks labelled with ``mode'' only have upper uncertainties, with the lower limit consistent with 0, and the maximum likelihood aspect ratio is labelled with dotted orange lines. }
    \label{fig:h}
\end{figure*}

The resulting constraints on the vertical aspect ratio of each disk are shown in Fig.~\ref{fig:h}. Depending on the shape of the PDF, each disk belongs to one of two classes: those with both upper and lower constraints, and those with only an upper limit. We defined a disk as having both an upper and lower constraint if the PDF drops to lower than half the peak on both sides of the peak. For these disks, we used the 16th, 50th and 84th percentiles to estimate the best-fit height and its uncertainties. These PDFs are relatively symmetric such that the median largely coincides with the maximum likelihood aspect ratio (or the ``mode''). For disks with only an upper limit, we estimated the best-fit height with the maximum likelihood aspect ratio rather than the median to avoid being significantly skewed by the asymmetric distribution. The upper uncertainties were estimated using the 84th percentile, whereas the lower uncertainties were simply set to extend to 0. These best-fit aspect ratios are the height assumptions for the radial profiles in the \texttt{fave} group in Fig.~\ref{fig:rp}. 

For the edge-on disks in the \texttt{rave} group, we also fitted their height as a function of radius using the non-parametric height fitting method in the original \texttt{rave} algorithm \citep{Han2022}. This approach recovers the height at different radii by iteratively adjusting the height of each annulus until the right amount of flux is concentrated along the edge-on disk's midplane. The number of annuli used for height fitting, $N_H$, and the distance from the major axis defining the midplane flux profile, $y_\text{mid}$, are shown in Table~\ref{tab:targets}. 

\begin{figure*}
    \centering
    \includegraphics[width=18cm]{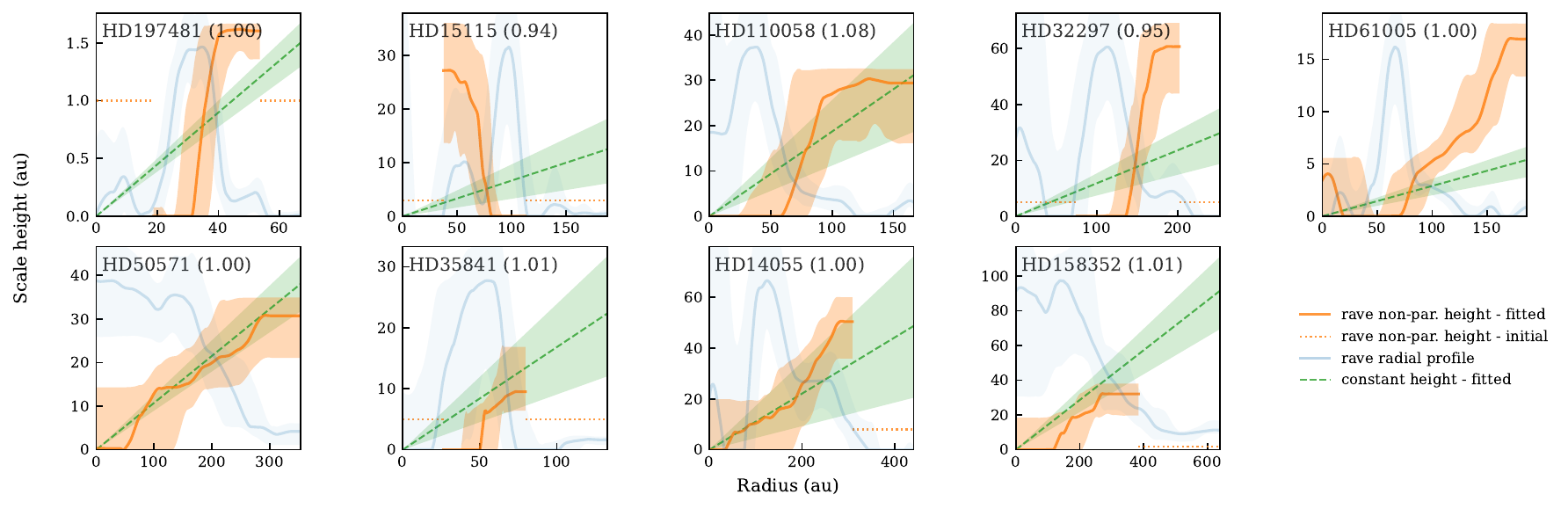}
    \caption{Comparison between radial scale height profiles of disks in the \texttt{rave} group fitted non-parametrically, shown in orange with solid lines, and best-fit heights assuming a constant aspect ratio, shown in green with dashed lines. The 1$\sigma$ uncertainties of the fitted aspect ratios about the median of the posterior distribution are indicated with shaded green regions. The radial profiles identical to those in Fig.~\ref{fig:rp} are over-plotted in blue to indicate regions with substantial emission. The left axis of each panel indicates the surface brightness (Jy/arcsec$^2$) and the right axis the scale height (au). The number in parentheses in each panel in the \texttt{rave} group is the ratio between the squared residuals of the non-parametric height profile model and the constant aspect ratio model. 
    For the \texttt{rave} heights as a function of radius, the dotted orange lines indicate the initial heights assumed during the iterative fitting procedure, and are plotted over the radial range over which there is negligible flux and where the height was not fitted. }
    \label{fig:rphp}
\end{figure*}

The non-parametric height distributions across radius are shown in Fig.~\ref{fig:rphp}, and the corresponding best-fit heights in the constant aspect ratio models are plotted for comparison. As the height is undefined in regions with negligible emission in the radial profile, the radial profiles are also shown to indicate the regions with substantial disk emission. We find that for HD\,14055, HD\,50571 and HD\,35841, there is a relatively close match between the shape of the $H(r)$ profile and the constant $h$ profile. For the remaining disks, the $H(r)$ profile is generally fitted to increase with radius (except for one system, HD\,15115). For some disks such as HD\,110058 and HD\,61005, the radial profiles appear to suggest a halo-like structure extending to large radii, which is also where the scale height appears to rapidly increase to a much larger value than in the main belt just interior to the halo. However, noting that $H(r)$ fitting can be significantly biased by noise in the image, these findings should be taken with caution given the limited resolution and sensitivity of these observations. 

Finally, a few diagnostics for the modelling are provided in Appendix~\ref{appendix}. Fig.~\ref{fig:model} and Fig.~\ref{fig:2d} display the best-fit model images and residuals, and Fig.~\ref{fig:1d} displays the 1D flux profiles that the algorithm fits to. The lack of significant residual structures suggest that the models fitted with \texttt{rave} and \texttt{fave} could adequately reproduce the disk structures within noise.

\subsection{Disk features}
\label{sec:diskfeatures}


The recovered radial profiles in Fig.~\ref{fig:rp} reveal a range of disk features across the sample. HD\,92945 \citep{Marino2019} and HD\,107146 \citep{Marino2018} are unambiguously shown to exhibit at least one radial gap in each system, as is also readily visible from the image and discussed in the corresponding references. HD\,197481 (AU~Mic, \citealp{Daley2019, Han2022, Marino2021}) and HD\,15115 \citep{MacGregor2013} show tentative evidence of a gap separating a secondary component interior to the main belt, consistent with the \texttt{frank} fits and previous studies in the corresponding references as discussed in Section~5.2.1 in \citet{Terrill2023}. Since the uncertainty regions also contain the possibility that such an inner component does not exist, we cannot conclude with certainty that these features exist, which \citet{Terrill2023} also warns against in their fit for HD\,15115. 

We also find evidence for outer sub-structures analogous to a halo, where the radial profiles decrease more slowly with radius than in the main belt. Such disks appear to include HD\,110058, HD\,32297, HD\,61005, HD\,9672 and HD\,10647. A few other systems, including AU~Mic, HD\,15115, HD\,109573, HD\,161868 and GJ\,14 also show very tentative signs of potential emission beyond the main belt, but given the level of uncertainty, it is difficult to discern whether these features are consistent with a halo or an additional belt separated by a gap, or whether these features are real in the first place. 

Indeed, even for the disks with a more significant detection of outer sub-structure, it could be difficult to discern between a halo and outer ring at the resolution and sensitivity of these observations. For HD\,61005, for example, \citet{Terrill2023} describes the emission recovered exterior to the main belt as potentially a secondary peak, whereas \texttt{rave} recovers a profile more consistent with a halo, which is consistent with the description in \citet{MacGregor2018}. The azimuthally averaged profile (Fig.~\ref{fig:1d}) is also well-reproduced by the profile with just a halo, without requiring an additional belt. The only halo systems noted in \citet{Terrill2023} are HD\,10647 and HD\,9672. Here, we listed more systems that could tentatively host halos, which will require deeper observations to confirm, including HD\,61005. 

Finally, for the remaining systems, including HD\,35841, HD\,14055, HD\,76582, HD\,191089, HD\,50571 and HD\,158352, we find no evidence for sub-structures beyond a main belt at the resolution and sensitivity of these observations. 

\subsection{Comparison between \texttt{rave}/\texttt{fave} and \texttt{frank}}
\label{sec:ravefrank}

With \texttt{rave} and \texttt{frank} both being non-parametric disk modelling tools, here we compare the results derived from the two methods. Note that the clean images we fitted to were not primary-beam corrected to ensure that the \texttt{rave} and \texttt{frank} profiles are comparable, as the primary beam correction was not applied when the \texttt{frank} profiles were fitted in \citet{Terrill2023}. In practice, the disks considered here are generally small compared to the primary beam so whether or not the primary beam correction is applied is not expected to make a significant difference. 

Fig.~\ref{fig:rp} overplots a version of the radial profiles derived by \texttt{frank} \citep{Terrill2023} on the \texttt{rave} profiles derived in this study. 
Comparison with the observed (and beam-convolved) azimuthally averaged profiles indicate that both methods have recovered sharper features than those that are directly visible from the observations. 
The main features recovered in both methods are overall highly consistent. However, we note three subtle differences between the two methods. 

Firstly, it appears that in general \texttt{frank} tends to return radial profiles that are smoothed to a certain characteristic length scale set by the regularisation, whereas \texttt{rave} tends to produce a more versatile range of shapes. This difference reflects in part the way in which the two algorithms attempt to reproduce the observations. The approach in \texttt{frank} involves fitting the radial profile directly to the visibilities, which is regularised by a Gaussian process \citep{Jennings2020}. The regularisation process effectively smooths the radial profile as the model visibilities are tapered beyond baselines where the S/N is too low. Attempting to produce extremely sharp features which deviate significantly from a sinusoidal-like shape could introduce oscillations due to the dampening of long baselines by the regularisation, such as those on either side of the main belt in HD\,109573 and along the outer edge of HD\,10647, which may not be physical. 
Similar ``bumps'' also exist on the outer edge of HD\,110058, HD\,32297 and HD\,61005, which may or may not be real features for this reason. While \texttt{rave} suffers from the same problem, it does not appear as pronounced as for \texttt{frank} in this dataset. For example, changes in slope on the outer edges of HD\,110058 and HD\,10647 could proceed without needing to invoke a change in concavity in the \texttt{rave} profiles. It is worth noting that hyperparameters in \texttt{frank} can be varied to mitigate oscillatory artefacts at the expense of spatial resolution \citep{Jennings2020}.

This difference between the two approaches can be both an advantage and disadvantage for each: recovering smoother shapes in the radial profile can make the fit more robust against noise while reproducing the large-scale structure of the disk, or be able to more sensitively pick out potential signs of multiple belts in the data. However, this could also introduce unphysical oscillations that appear like multiple belts when, in reality, these features may be consistent with sharp changes in slope. This highlights the benefit of using both approaches to combine the strengths of each to reach better informed conclusions. We suggest that future work could test both \texttt{rave} and \texttt{frank} on simulated observations with known radial profiles, particularly those with sharp changes in slope, such as low-amplitude halo-like features in HD\,61005, to more unambiguously benchmark the subtle differences in performance between the two approaches for these types of radial profiles. 

A second subtle difference between the two methods is that the version of \texttt{frank} used to present radial profiles in \citet{Terrill2023} was permitted to take negative values---although the \texttt{frank} option to recover radial profiles without negative values is available, which could also help to remove many of the oscillatory artefacts---whereas \texttt{rave} forces the radial profiles to be non-negative. In both methods, direct fits to the observations often inevitably return negative values due to a combination of noise in the observations and oscillatory behaviour intrinsic to the methods. If any negative values exist in the fitted radial profile, \texttt{rave} sets these values to 0 before outputting the final profile. This may decrease the goodness of fit, but it is the expense for obtaining a radial profile without unphysical negative values. Comparison between the data and fitted model for \texttt{rave} (Figs.~\ref{fig:2d} and \ref{fig:1d}) suggests that flooring the radial profile to 0 does not in general jeopardise the quality of fit in this dataset. 

Finally, the uncertainty range estimated by \texttt{rave} is more conservative than that estimated by \texttt{frank}. As Section~\ref{sec:fave-radial} describes, the interpretation of the uncertainty range in \texttt{rave} differs from parametric models in that they are not uncertainties conditioned on a specific model, but instead indicate the range of different non-parametric models which could all plausibly be consistent with the observations. While the radial profiles of disks with gaps including HD\,92945 and HD\,107146 are relatively tightly constrained, the range of possible models in other systems could be important for interpreting whether any sub-structures are real. For example, for AU~Mic (Fig.~\ref{fig:rp}), the inner peak is highly tentative according to \texttt{rave} since the range of possible models is almost consistent with the inner peak being absent. Similarly, for HD\,32297 and HD\,191089, the large uncertainties suggest that the innermost region is very poorly constrained. In practice, however, the uncertainties in this region may be somewhat overestimated since their images suggest the emission near the star to be likely at a low level. For \texttt{frank}, we observe that the 1$\sigma$ range is significantly smaller, including in systems with oscillatory behaviour, such as HD\,109573 and along the outer edge of HD\,10647. The fact that the uncertainty regions around these oscillations are small in the \texttt{frank} profiles may appear to disfavour a scenario without such oscillations, however if we were to consider that these features may not be real, these uncertainties may be somewhat underestimated. Overall, we highlight that while the \texttt{rave} and \texttt{frank} uncertainties carry different interpretations and are not directly comparable, their combined information could help elucidate the significance of any sub-structures detected. 

\begin{figure}
    \centering
    \includegraphics[width=8cm]{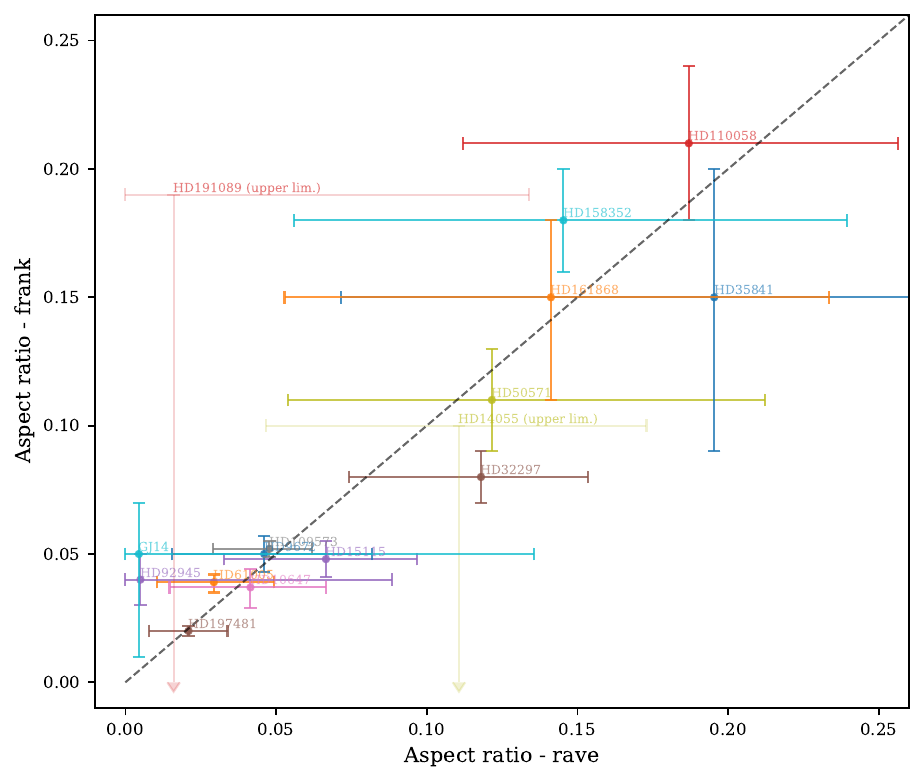}
    \caption{Comparison between the scale heights in the sample fitted with \texttt{rave}/\texttt{fave} in this study and with \texttt{frank} in \citet{Terrill2023}. The dashed diagonal line has a slope of 1 and should intersect the uncertainties of each point if the \texttt{rave} and \texttt{frank} estimates are consistent. Note that HD\,191089 and HD\,14055 plotted in lighter shades only show the upper limit reported for \texttt{frank}. }
    \label{fig:h_comparison}
\end{figure}

To compare the vertical height fitting results, we plotted the heights fitted by \texttt{frank} from \citet{Terrill2023} against those fitted by \texttt{rave} in Fig.~\ref{fig:h_comparison}. For disks with both upper and lower constraints on the scale height fitted with \texttt{rave} (labelled as ``median'' in Fig.~\ref{fig:h}), we plotted the aspect ratio value and uncertainties using the 16th, 50th and 84th percentiles of the PDF. For the disks without a lower limit on the scale height (labelled as ``mode'' in Fig.~\ref{fig:h}), we plotted the centre of the point at the maximum likelihood scale aspect ratio, setting the upper uncertainties based on the 84th percentile and plotting the lower uncertainties such that they drop towards 0. The \texttt{frank} uncertainties correspond to the 1$\sigma$ estimate where it is provided in \citet{Terrill2023}, and only the upper limit is shown for the two disks without lower limits. 

Overall, we find that the best-fitting estimates between the two methods are in close agreement. For two systems, HD\,92945 and GJ\,14, the \texttt{rave} best-fit estimate (closer to 0) is significantly lower than the \texttt{frank} value (closer to 0.04 and 0.05), but the two still agree within uncertainties given the large uncertainty margins on the \texttt{rave} values. In general, \texttt{rave} returns larger uncertainties on the vertical height than \texttt{frank}. As a result, more disks fitted with \texttt{rave} do not provide tight constraints on the lower limit of the scale height, as reflected in the uncertainty bars that drop towards 0 in Fig.~\ref{fig:h_comparison}. The difference in uncertainties in fitted scale heights could be related to how the observational uncertainties are estimated: for \texttt{frank} these are directly based on entries in the visibility table, whereas for \texttt{rave} these are based on background noise measurements integrated over the area of disk emission. Assuming a $\chi^2$ distribution in image space may not fully take into account any correlation of the noise between different spatial regions across the image, thereby over-estimating the noise and uncertainties, whereas the noise in visibility space may be more reflective of the uncertainties directly placed on the type of data collected. However, as the examples of HD\,92945 and GJ\,14 show, the relatively large uncertainties for \texttt{rave} may be justified given that the results from the two methods could otherwise become inconsistent. 

\subsection{Stellar and disk properties}
\label{sec:correlations}
Our uniform analysis of the three-dimensional structure in an inclined sample of disks provides an opportunity to explore the stellar and disk properties across the sample and any correlations between them. We derive and summarise these system properties in the following sections.

\subsubsection{System properties}
\label{sec:basic_properties}

\begin{figure*}
    \centering
    \includegraphics[width=18cm]{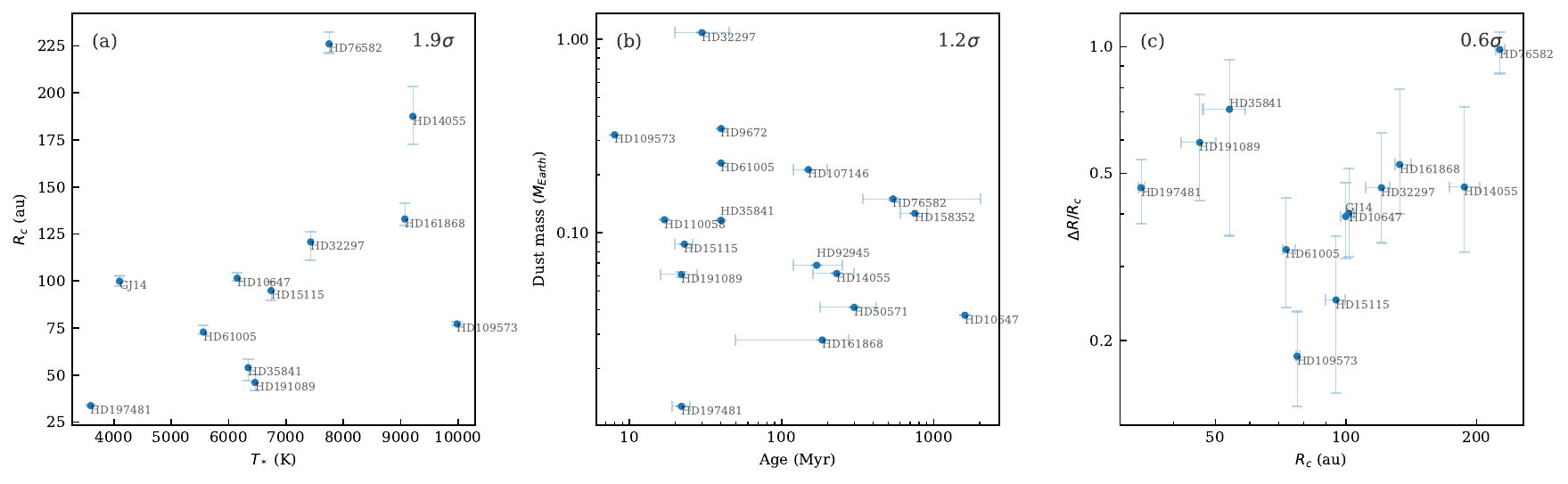}
    \caption{Summary of basic stellar and disk properties in the sample, including the centroid radius ($R_c$), FWHM of the radial profile ($\Delta R$), stellar temperature ($T_*$), dust mass and age. Note that panel (b) contains points without uncertainty estimates in age (see Fig.~\ref{tab:targets}). The significance of a linear correlation within each panel is displayed on the upper-right corner. The brown dotted line in panel (c) indicates the line of best fit. }
    \label{fig:basic_properties}
\end{figure*}

To explore any trends relating to disk structure, we characterised disks within the sample with a few parameters of interest. 
\begin{itemize}
    \item Centroid radius ($R_c$): the mean radius weighted by surface brightness. We estimated the uncertainties using the 1$\sigma$ interval of $R_c$ calculated from 100 individual radial profile fits with randomised annuli boundaries (Section~\ref{sec:fave-radial}), which are the same individual fits used to estimate the range of possible models of the radial profile. 
    
    \item Inner edge width: the number of au's required for the radial profile of the main belt to rise from 25\% to 75\% of the peak value. This parameter is a proxy for the inverse of the inner edge slope. To estimate the uncertainties, we calculated the extremum of the inner edge width using the extreme values within the range of possible models (i.e., using the smallest radius that falls within the range of possible models at 25\% of the peak value, and the largest radius at 75\%, to estimate the shallowest slope, and vice versa for the steepest slope). The range of possible models that we fitted encapsulate a range of different profile shapes with complex correlations across radius, which are different from uncertainties on a specific functional form in parametric fitting. We estimated the upper and lower uncertainties on the edge width as half the difference between that of the median profile and those of the 2 extremes, as it is less likely for a radial profile to take opposite extremes at both 25\% and 75\% of the peak than at only one point.
    
    \item Outer edge slope: the number of au's required for the radial profile of the main belt to fall from 75\% to 25\% of the peak value. The uncertainties were estimated using the same approach as for the inner edge slope, but applied to the outer edge. 
    
    \item FWHM of the radial profile ($\Delta R$): measured from the median radial profiles shown in Fig.~\ref{fig:rp}. We estimated the uncertainties using an analogous approach as for the edge widths, computing the maximum and minimum FWHM within the range of possible models, and estimating the upper and lower uncertainties as half the difference between the FWHM of the median profile and those of the 2 extremes. 
    
    \item Vertical height aspect ratio ($h$): same as those fitted in Section~\ref{sec:modelling}. We used the constant aspect ratio models as they have the fewest degrees of freedom and are likely to be a more robust reflection of the overall scale height. Uncertainty bars will be subsequently plotted to reflect the PDFs displayed in Fig.~\ref{fig:h}. The best-fit values and uncertainties are the median and 16th and 84th percentiles of the PDF for those labelled with ``median'', and the mode and 0th and 84th percentiles for those labelled with ``mode'' whose lower bounds are consistent with 0. 
    
    \item Dust mass: the sub-millimeter dust mass in the disk estimated with
    \begin{equation}
        M_\text{dust} = 4.25 \times 10^{10} \frac{1}{\kappa_\nu} \int_0^{\infty} \frac{f(r)}{B_\nu(\lambda, T(r))} 2 \pi r dr,
    \end{equation}
    where $f(r)$ is the face-on surface brightness profile in Jy/arcsec$^2$, $r$ is the radius in au, $B_\nu$ is the Planck function in Jy/sr and $\kappa_\nu$ is the mass opacity assumed to be 45~au$^2$/M$_\text{Earth}$ \citep{Beckwith2000, Wyatt2008}. The temperature profile is assumed to be
    \begin{equation}
        T = 278.3 \, r^{-1/2} L_*^{1/4},
    \end{equation}
    where $L_*$ is the stellar luminosity in solar luminosities, $r$ is in au and $T$ is in Kelvins \citep{Wyatt2008}. Note that this approximation assumes a constant mass opacity across all systems. The uncertainties thus only carry those from the radial profiles, and are estimated using the 1$\sigma$ interval from 100 independent fits with randomised annuli boundaries, analogous to the uncertainties of the centroid radius. 
\end{itemize}

Note that the inner edge width, outer edge width and $\Delta R$ are only well defined if there is only one clear belt within the radial profile. When calculating these 3 quantities, we therefore excluded disks with clear multi-belt structures with gaps (HD\,92945 and HD\,107146) and those whose inner edges do not drop to a value close to 0 (HD\,110058, HD\,50571, HD\,158352 and HD\,9672, possibly due to resolution and sensitivity limitations). 

In addition to measuring these disk quantities, we also collected the stellar spectral type, stellar temperature, $T_*$, from SED modelling (G. Kennedy, private communication) and age of each system, as displayed in Table~\ref{tab:targets}. 

Some basic properties of the sample are plotted in Fig.~\ref{fig:basic_properties}, which exhibit expected trends that have been explored in previous studies. Panel (a) shows an increasing centroid radius with stellar temperature. This behaviour across the sample is consistent with population analyses of disks resolved in the millimeter \citep{Matra2018} and tentatively in the far-infrared \citep{Pawellek2014}. It is also consistent with dust temperature modelling of disk populations that show either a common dust temperature across cold debris belts \citep{Morales2011} or a dust temperature that increases with stellar temperature \citep{Ballering2013}. Although the former is more consistent with ice lines playing a major role in setting belt locations \citep{Matra2018}, and the latter suggestive of the importance of additional factors \citep{Hughes2018}, either scenario in dust temperature is consistent with an increasing dust belt radius as a function of stellar temperature. 

Panel (b) in Fig.~\ref{fig:basic_properties} shows an overall decreasing dust mass with age, albeit with a large scatter and low significance. This is consistent with findings from the SONS survey with the James Clerk Maxwell Telescope (JCMT) at sub-millimeter wavelengths \citep{Holland2017} and from the DUNES and DEBRIS surveys with \textit{Herschel} in the far-infrared \citep{Montesinos2016}, and could be reflective of collisional erosion of solid disk material over time \citep{Wyatt2008, Hughes2018}.

Panel (c) in Fig.~\ref{fig:basic_properties} shows that the fractional radial widths of disks in the sample span a large range of values, which has previously been known to exist in debris disks \citep{Hughes2018}. These $\Delta R / R_c$ values exhibit no clear correlation with $R_c$. The fractional width is expected to be constant if the semi-major axis distributions are roughly ``scaled up'' according to disk radius---which in turn is partially set by the stellar luminosity (panel a)---and the eccentricity dispersion is comparable between disks. The wide range of fractional widths observed could therefore reflect a diverse range of semi-major axis and/or eccentricity distributions among debris disks. 
We discuss radial and vertical characteristics across the sample in more detail in the following sections.

\begin{figure*}
    \centering
    \includegraphics[width=18cm]{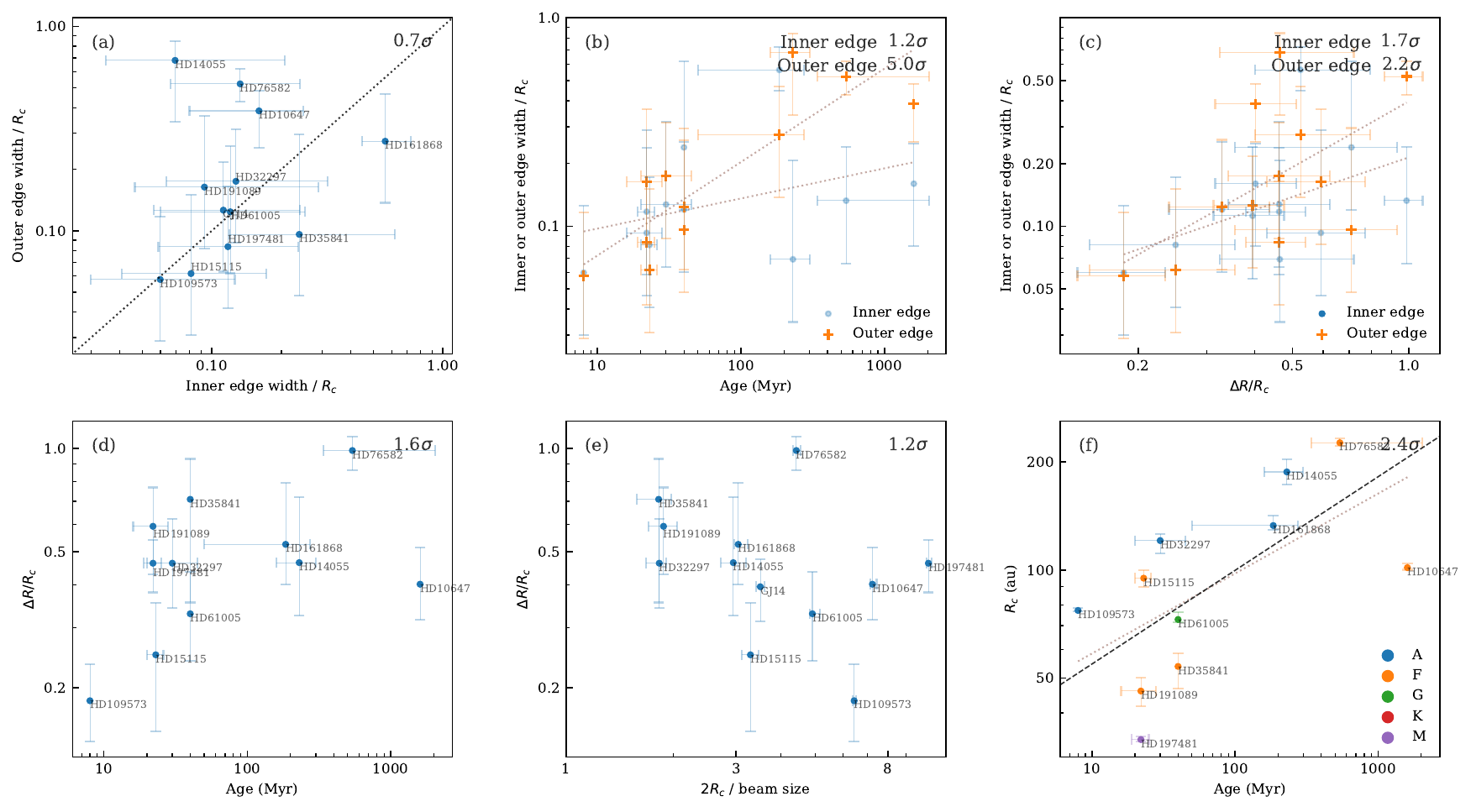}
    \caption{Summary of radial properties in the sample. In addition to the centroid radius ($R_c$) and FWHM of the radial profile ($\Delta R$) shown in Fig.~\ref{fig:basic_properties}, the inner and outer edge widths are also shown, which represent the number of au's over which the radial profile rises/falls from 25\% to 75\% of the peak surface brightness/vice versa. The black dotted line in panel (a) indicates points where the fractional inner and outer edge widths are equal. In panel (f), points are coloured by spectral type and the dashed line indicates $R_c = 30\,\text{au} \, (\text{Age} / \text{Myr})^{6/23}$. The significance of a linear correlation within each panel is displayed on the upper-right corner. Brown dotted lines indicate lines of best fit, where the inner and outer edges are fitted separately in panels (b) and (c). }
    \label{fig:radial_analysis}
\end{figure*}

\begin{figure*}
    \centering
    \includegraphics[width=18cm]{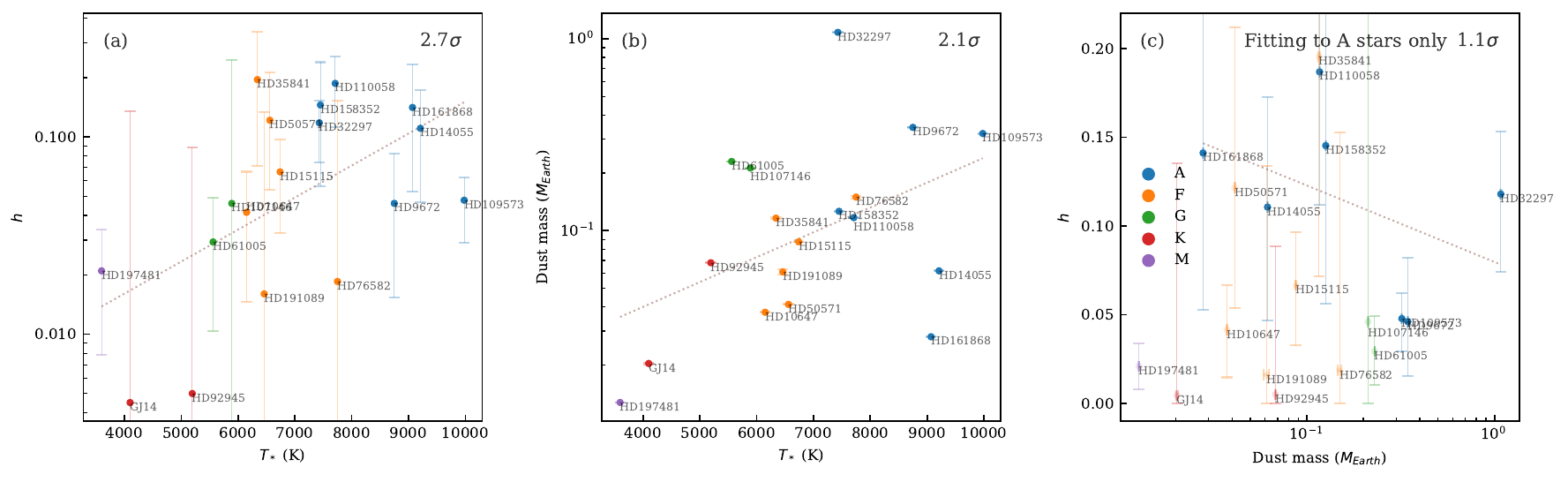}
    \caption{Summary of the vertical height aspect ratio ($h$), stellar temperature ($T_*$) and dust mass in the sample. Data points are coloured by spectral type. A-type stars which span the largest range in estimated dust mass are emphasised in panel (c) in an attempt to isolate the effects of spectral type and dust mass. The significance of a linear correlation within each panel is displayed on the upper-right corner. Brown dotted lines indicate lines of best fit. }
    \label{fig:height_analysis}
\end{figure*}

To provide some degree of quantification of the strength of any potential correlations between stellar and disk properties in Figs.~\ref{fig:basic_properties}--\ref{fig:no_correlation} that will be subsequently discussed, we uniformly estimated the significance of a linear correlation between each pair of quantities by dividing the absolute value of the least-squares linear slope, fitted in either linear or log space as indicated by the axes of each panel, by the standard error of the slope. The estimated significance values are displayed on the upper-right corner of each panel in these figures. 

\subsubsection{Radial structure}
\label{sec:radial_correlations}

In this section, we explore several trends related to the radial structure seen across the sample, which are summarised in Fig.~\ref{fig:radial_analysis}.

The non-parametric nature of our modelling offers an opportunity to examine debris disk features such as the degree of symmetry between the inner and outer edges. A direct comparison between the fractional inner and outer edge widths in panel (a) shows an equal number of systems with a fractionally wider inner edge compared to those with a fractionally wider outer edge. These measurements show that while the radial profiles within this sample are on average largely symmetric, there is a large scatter in the direction and extent to which the belts are skewed. There also appears to be no strong correlation between the fractional inner and outer edge widths, reflecting a diverse range of radial profile shapes that exist in this sample. 

These disk edge measurements are based on the slope between 25\% and 75\% of the peak of the main belt, but could also be affected by the presence of inner or outer halos. 
Closer inspection of the sample suggests that the 4 disks above 100~Myr are also those with the widest outer edge in both fractional and au terms, 3 of which have a wider outer edge than inner edge. For HD\,10647, this could reflect the presence of the outer halo, whereas for HD\,76582 and HD\,14055, the main belt itself appears to be skewed, although the distinction between a halo and a skewed belt is not necessarily a clear one. 

Panel (b) displays the fractional inner and outer edge widths against age. Interestingly, the fractional outer edge width appears to correlate positively with age (5.0$\sigma$). If real, this could suggest that more evolved disks more likely exhibit shallower outer edges. Such an effect could occur if mechanisms such as planet migration are at play, which could create a scattered population of planetesimals on eccentric orbits (and randomised in pericenter alignment) that simultaneously increases the width of the main belt and the width of the outer edge (e.g., \citealp{Marino2021}), analogous to the formation of a scattered disk in the Solar System due to the influence of Neptune and its migration \citep{Morbidelli2004}. Further investigation is required to understand whether such a broadening would be significant enough to explain these observations.

The fractional outer edge width also tentatively (2.2$\sigma$) correlates positively with $\Delta R / R_c$, as shown in panel (c). It is therefore expected that $\Delta R / R_c$ should also increase with age, which panel (d) displays and confirms at a low significance. Mechanisms that broaden the disk, such as by increasing the eccentricity dispersion, could conceivably affect all of its radial features. In general agreement with this expectation, the fractional inner edge width also appears to increase with both age and $\Delta R / R_c$, as shown in panels (b) and (c), however at a much lower significance, possibly reflecting a more complex range of planetary configurations and dynamical effects at play closer to the inner disk (e.g., spreading of the disk's inner edge could be limited by a planetary system which removes debris that migrates inwards). 

To investigate whether the correlation between the fractional FWHM and inner and outer edge widths could be caused by biases in the fitted radial profiles, which could be due to differences in the resolution used to observe these disks, panel (e) shows the fractional disk width as a function of the number of resolution elements across the disk's major axis. There does not appear to be a significant correlation between the two, suggesting that the disks found to have a small fractional width are not necessarily better resolved than those with large fractional widths measured. The same lack of significant correlation with resolution applies to the fractional inner and outer edge widths. This supports the interpretation that the correlations between disk width and inner and outer edge slopes are more likely to be real than due to limitations in resolution. 

While the correlations discussed so far have focused on fractional terms, i.e., adjusted for the centroid radius, we find tentatively (2.4$\sigma$) that the centroid radius itself appears to increase with age, as shown in panel (f). In models of self-stirred disks, the radius of peak dust emission is expected to increase over time as the disk collisionally evolves, consistent with what we observe in this sample. The peak radius set by the collisional time is expected to scale as time to the power of 6/23 \citep{Kennedy2010}. An indicative curve showing the slope of this scaling is plotted in panel (d). However, other mechanisms could also delay stirring in the outer regions of the disk, such as planet stirring via secular perturbation. It is difficult to conclude whether the centroid radius and age correlation in this limited sample truly reflects a stirring mechanism in these systems. Alternatively, it is possible that disks that started out with larger masses also have larger radii, and those that are still observable after collisionally evolving for longer periods of time correspond to those that started out being radially large \citep{Wyatt2007}. 


Finally, we find no correlation between the dust mass and the three disk width parameters---the fractional radial FWHM and fractional inner and outer edge widths. We also find no correlation between the stellar temperature and the three disk width parameters. 

\subsubsection{Vertical height}
\label{sec:vertical_correlations}

In this section, we explore several trends related to the scale height across the sample. Fig.~\ref{fig:height_analysis} summarises the sample with two-dimensional slices in the parameter space of vertical height aspect ratio, stellar temperature and dust mass. 

We find that the vertical height aspect ratio appears to broadly increase (2.7$\sigma$) towards larger stellar temperatures, as shown in Fig.~\ref{fig:height_analysis} panel (a). The uncertainties here indicate the PDFs of the vertical height aspect ratio derived in Fig.~\ref{fig:h} and described in Section~\ref{sec:basic_properties}, which translates to large uncertainties on this plot. However, noting that the \texttt{frank} heights fitted independently by \citet{Terrill2023} are in close agreement with our best-fit height measurements indicated with round points (except for the 2 K stars, Fig.~\ref{fig:h_comparison}), the trend suggested by the best-fit heights could be more robust than the \texttt{rave} height uncertainties suggest. Substituting the \texttt{rave} heights with those from \texttt{frank} produces the same apparent trend while having smaller height uncertainties derived directly from uncertainties in the visibility measurements, as discussed in Section~\ref{sec:ravefrank}. 

\begin{figure*}
    \centering
    \includegraphics[width=18cm]{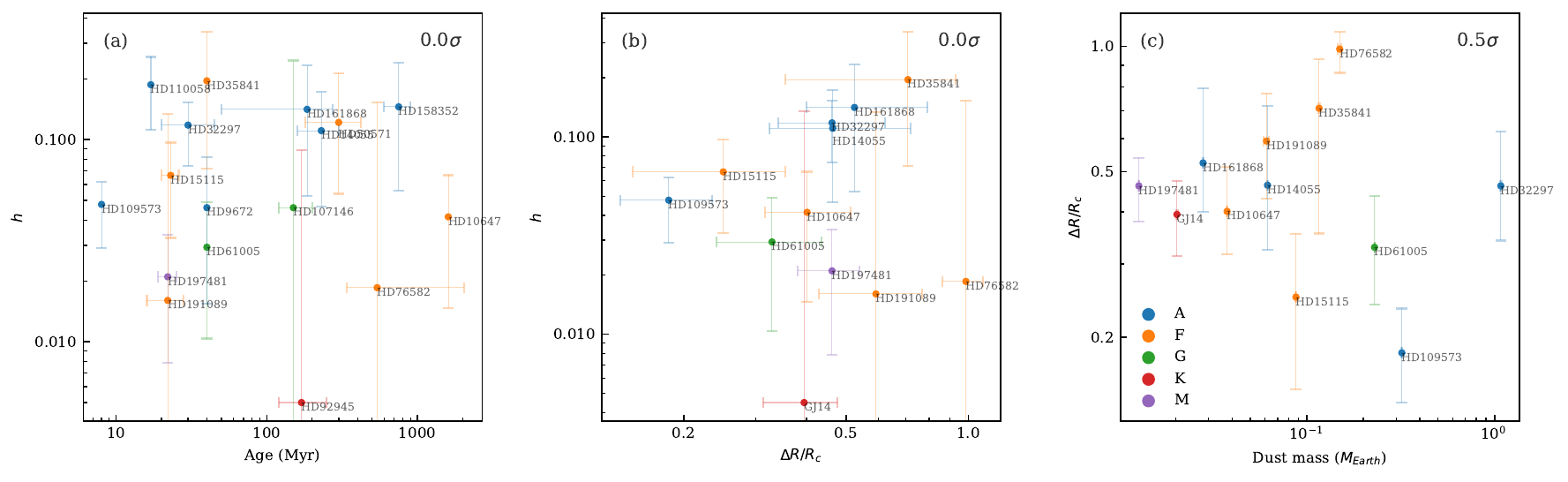}
    \caption{Summary of stellar and disk properties which do not show clear or monotonic correlations with each other. The significance of a linear correlation within each panel is displayed on the upper-right corner.}
    \label{fig:no_correlation}
\end{figure*}

To better understand the tentative correlation between scale height and stellar temperature, we note that the scale height could be influenced by the mass of the disk, as a larger mass in the planetesimal belt could either increase large stirring bodies and hence the vertical height or damp the vertical dispersion in the presence of inclined planets due to the disk's self-gravity \citep{Poblete2023}. As Fig.~\ref{fig:height_analysis} panel (b) suggests, a larger dust mass in the disk is indeed tentatively seen around more massive stars, potentially reflecting a larger planetesimal mass reservoir in the high-mass end of the collisional cascade. 
Panel (c) plots the vertical height aspect ratio against dust mass, which shows a broadly positive correlation, although the scatter is large. In an attempt to isolate the effect of stellar mass or temperature from the dust mass on the vertical height, we emphasised the A stars plotted in blue in panel (c), which span the largest range in dust mass. Although a decreasing aspect ratio with dust mass is possible among the A stars, no clear correlation can be confidently concluded given the large scatter.

Interestingly, we do not see a clear correlation between the scale height and age, as shown in Fig.~\ref{fig:no_correlation} panel (a). However, this does not necessarily rule out theoretical predictions that the inclination dispersion should increase with time, such as the prediction that $h \propto \text{age}^{1/4}$ by \citet{Quillen2007}, which would only result in a factor of 4 increase over the age range probed by this sample. 

We also do not see a clear trend between the fractional radial width and the vertical aspect ratio, as shown in Fig.~\ref{fig:no_correlation} panel (b), although a larger scatter or divergent populations in height towards larger fractional widths is possible. Note that the distribution in panel (b) differs from those in \citet{Terrill2023}, who tentatively found a bimodal distribution in the vertical height aspect ratio and that the aspect ratio tentatively correlates with fractional width. This difference is in part due to the fact that of the 5 disks measured to have a fractional width of above 1 in \citet{Terrill2023}, 3 were excluded from width measurements in this study (HD\,158352, HD\,110058 and HD\,9672) as the inner edges of their radial profiles are not well-resolved (Fig.~\ref{fig:rp}), and the remaining 2 (HD\,35841 and HD\,14055) were measured to have a smaller radial FWHM in this study. 

Similarly, no clear correlation is seen between the fractional radial FWHM and dust mass in Fig.~\ref{fig:no_correlation} panel (c), but there appears to be a larger scatter in fractional widths at larger dust masses among the limited number of systems in the sample.
If the divergent aspect ratio as a function of radial fraction width in panel (b) and the divergent fractional width as a function of dust mass in panel (c) are real, this may suggest that larger disk masses could enable a wider range of formation and evolutionary outcomes in terms of the fractional disk width, the broader ones of which could in turn enable different pathways that result in a wider range of scale heights. 

Finally, the three systems with gas detections, HD\,9672, HD\,110058 and HD\,32297, and those without do not suggest a correlation between the presence of gas and the disk's vertical height in this sample. Gas detection has not been found to correlate strongly with dust mass \citep{LiemanSifry2016}, so the lack of correlation between gas and dust properties found in this limited sample in the millimeter may not be surprising. We suggest future work with vertically resolved disk imaging to further populate and test this and other potential trends in Fig.~\ref{fig:no_correlation}.

\subsubsection{A speculative story}

Though our sample is limited in size, it is unique in providing an opportunity to simultaneously study the radial and vertical structure with resolved millimeter imaging in a uniform analysis that mitigates against biases in fitting the radial profile. The comparison enabled by two different methods, \texttt{rave} in this study and \texttt{frank} in \citet{Terrill2023}, provides some degree of evidence for the robustness of the fitting. 


Based on these findings, we may speculate whether there could be a coherent story that links the general trends which we see in this sample, even if some remain tentative. Planet-sized bodies are thought to dominate the mass of the planetary system and their perturbations could significantly influence the evolution of debris disks. Among its range of dynamical outcomes could be an outer edge that becomes shallower with age (Fig.~\ref{fig:radial_analysis}b), which could occur via the scattering of planetesimals outwards into a scattered disk component. A scattered disk would also skew the shape of the radial profile, with the 4 oldest systems also being the fractionally broadest, 3 of which are skewed with a wider outer edge than inner edge or which exhibit an outer halo. 
Simultaneous to the fractional broadening of the outer edge over time, the centroid radius increases (Fig.~\ref{fig:radial_analysis}f, e.g., \citealp{Kennedy2010}), which could potentially be driven by the faster collision and depletion rates at smaller radii that cause an outwards propagation of the peak emission, or the case that only the more massive and extended disks remain detectable in older systems. The emergence of a scattered disk could further increase the centroid radius while making the disk appear fractionally broader (Fig.~\ref{fig:radial_analysis}c, d). 

Clues to the presence of the most massive bodies in the system could also have been left on the vertical structure, with the vertical height aspect ratio being larger around more massive stars (Fig.~\ref{fig:height_analysis}a) which are more likely to host larger stirring bodies, such as embedded planetesimals or inclined planets. The presence of more massive planetesimal belts around more massive stars could be suggested by the larger mm dust mass (Fig.~\ref{fig:height_analysis}b) produced from a collisional cascade. However, while a larger disk mass could imply the more likely presence of larger stirring bodies, a more massive planetesimal disk could also damp the effects of any inclined stirring planets (\citealp{Poblete2023}), making the overall relationship between dust mass and scale height less clear, although a very tentatively decreasing aspect ratio with dust mass is possible when controlling for stellar mass (Fig.~\ref{fig:height_analysis}c). 
Although the fractional radial widths appear to vary with age, this does not appear to be the case for the vertical aspect ratio (Fig.~\ref{fig:no_correlation}a). It is possible that quasi-equilibrium is reached vertically over shorter timescales and subsequently evolves more slowly compared to the steady radial evolution over time.
The potentially larger scatter in vertical aspect ratio in fractionally wider disks (Fig.~\ref{fig:no_correlation}b) could reflect a greater diversity in orbital configurations of any perturbing planets relative to broader disks and the range of disk masses which damp their effects. These hypothetical planets and the disk mass would appear to have weaker effects on the fractional radial width, so no clear correlation is observed between the fractional width and dust mass (Fig.~\ref{fig:no_correlation}c). 

Over time, the dust mass decreases as the disk collisionally erodes (Fig.~\ref{fig:basic_properties}b, \citealp{Holland2017}). 
Despite any broad structural trends, the large scatter in fractional widths (Fig.~\ref{fig:basic_properties}c) and inner and outer edge asymmetries (Fig.~\ref{fig:radial_analysis}a) while the disks are still observable record the dynamical interactions set off by wide ranging initial planet and disk configurations, whereas the correlation between belt radius and stellar luminosity (Fig.~\ref{fig:basic_properties}a) following many Myr to Gyr of evolution could still be hinting at their initial formation in the protoplanetary disk, possibly being partly set by the location of ice lines (e.g., \citealp{Morales2011, Matra2018}). 

We conclude by noting that given our sample predominantly comprises of the best resolved and brightest disks observed by ALMA, the sample could be biased towards more extended and bright disks, which may not reflect the full range of possible evolutionary scenarios of all debris disks. For example, \citet{Najita2022} proposed several possible scenarios through which debris rings could form from protoplanetary disks based on evolutionary models, some of which correspond to disks with lower masses which may be more difficult to detect than those in this study. Future observations that focus on fainter and more compact disks as well an expansion of the sample and its diversity will help elucidate possible scenarios describing the dynamical evolution of debris disks.

\section{Conclusions}
\label{sec:conclusions}
We summarise this paper in the following points. 

1. \citet{Han2022} developed a method to model the three-dimensional structure of axisymmetric, optically thin disks such as debris disks from thermal emission images, deriving their surface brightness and (if sufficiently edge-on) vertical height as a function of radius without having to assume their functional forms, returning realistic uncertainties independent of model assumptions. We further advanced this method in this paper, developing a modified alglorithm optimised for non-edge-on disks that performs deconvolution and de-projection at higher resolution and is more robust against noise. This new method still allows for the vertical height aspect ratio of the disk ($h = H(r)/r$) to be inferred. 

2. We further extended the applicability of this non-parametric approach to images involving anisotropic emission, such as light from a star scattered by a dust disk, by empirically deriving the scattering phase function from observations and taking it into account when performing the non-parametric fitting. 

3. We used simulated observations with known structures as test cases to demonstrate that the face-on optimisation can effectively de-project and deconvolve images of axisymmetric, optically thin disks with a range of radial profiles, and recovers radial profiles more robustly and accurately than the original algorithm. Scattered light fitting is effective if a high S/N is reached, even when information from a large range of azimuth is lost in the image, such as due to PSF subtraction artefacts or other imaging artefacts in the form of radial spikes. Low levels of artefacts among available scattered light observations are still required for the fitting to proceed robustly. 

4. While the face-on extension performs more effectively for non-edge-on disks, the original algorithm is the preferred approach for edge-on disks. Together, the methods developed allow for non-parametric recovery of disk structure from images at all wavelengths and across all inclinations. The extended algorithm, \texttt{fave}, is implemented as part of the \texttt{Python} package for the original algorithm optimised for edge-on disks, \texttt{rave} \citep{Han2022}. 

5. We applied \texttt{rave}/\texttt{fave} to recover the three-dimensional structure of a sample of 18 inclined debris disks resolved with ALMA, which is largely based on the sample studied by \citet{Terrill2023} who fitted the observations with a visibility-space de-projection and deconvolution algorithm, \texttt{frank} \citep{Jennings2020}. Comparison between the fitted radial profiles and vertical height between \texttt{rave}/\texttt{fave} and \texttt{frank} reveal consistent results, although subtle differences exist which are discussed in more detail in 
Section~\ref{sec:ravefrank}.


6. Among the disks in the sample with a single main belt, we find a wide range of fractional radial widths that fall between 0.2 and 1.0. The sharpness of the outer edge appears to decrease with age, which could reflect the outwards scattering of planetesimals by migrating planets over time, although more detailed modelling is required to test this scenario. The centroid radius increases with age in the sample, which appears broadly consistent with expectations of collisional evolution under either self-stirring or planet-induced secular perturbation, but observational biases that preferentially detect initially broader and more massive disks are possible. The fractional width of the disk and its edges do not appear to correlate with dust mass or stellar temperature. 

7. In contrast to the radial trends, the vertical height aspect ratio is tentatively found to increase with stellar temperature, but does not correlate in a clear way with age or the fractional disk width. These vertical aspect ratio trends could reflect an interplay between planet stirring and the disk's self gravity, both of which likely increases with stellar mass but produces opposite effects on the scale height.

\section*{Acknowledgements}
The authors are grateful for discussions with Julien Milli and Johan Olofsson on scattered light imaging, which provided inspiration for the further development of the methods presented here. Y.H. is funded by a Gates Cambridge Scholarship from the Gates Cambridge Trust enabled by Grant No. OPP1144 from the Bill and Melinda Gates Foundation. S.M. is supported by a Royal Society University Research Fellowship (URF-R1-221669). This research made use of NASA's Astrophysics Data System; the \textsc{IPython} package \citep{ipython}; \textsc{SciPy} \citep{scipy}; \textsc{NumPy} \citep{numpy}; \textsc{matplotlib} \citep{matplotlib}; and \textsc{Astropy}, a community-developed core Python package for Astronomy \citep{astropy}.

\section*{Data Availability}
The ALMA data used in this study are available on the ALMA Science Archive by querying targets in the sample.

\bibliographystyle{mnras}
\bibliography{references}

\appendix

\section{Appendix A}
\label{appendix}

\begin{figure*}
    \centering
    \includegraphics[width=18cm]{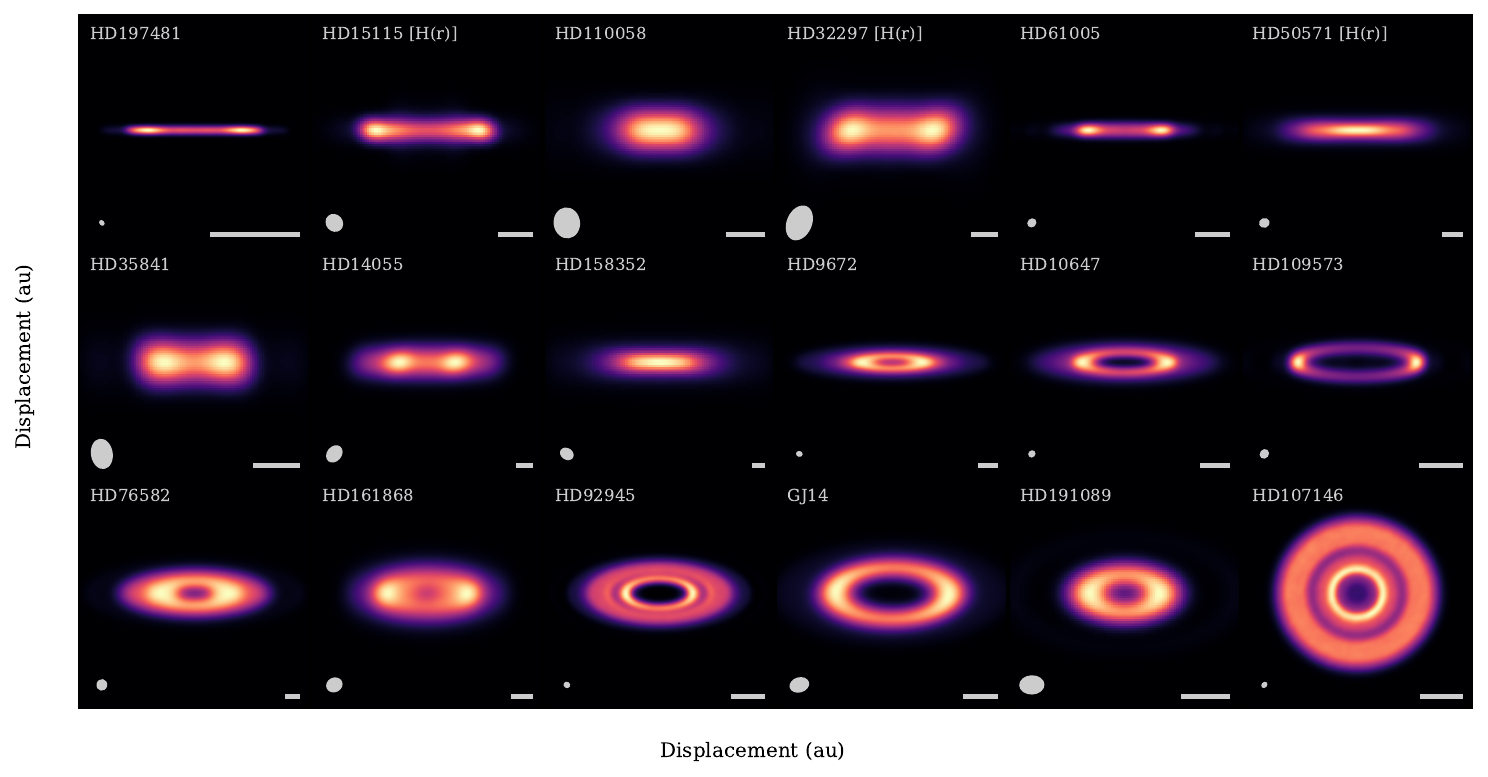}
    \caption{Model images of disks in the sample simulated using the radial profiles recovered with \texttt{rave} or \texttt{fave}. The vertical heights are assumed to be the best-fit aspect ratios shown in Fig.~\ref{fig:h}, except for those indicated with ``H(r)'' which use the non-parametric height profiles fitted with \texttt{rave} and can more closely reproduce the observations than the best-fit constant aspect ratio models. }
    \label{fig:model}
\end{figure*}

\begin{figure*}
    \centering
    \includegraphics[width=18cm]{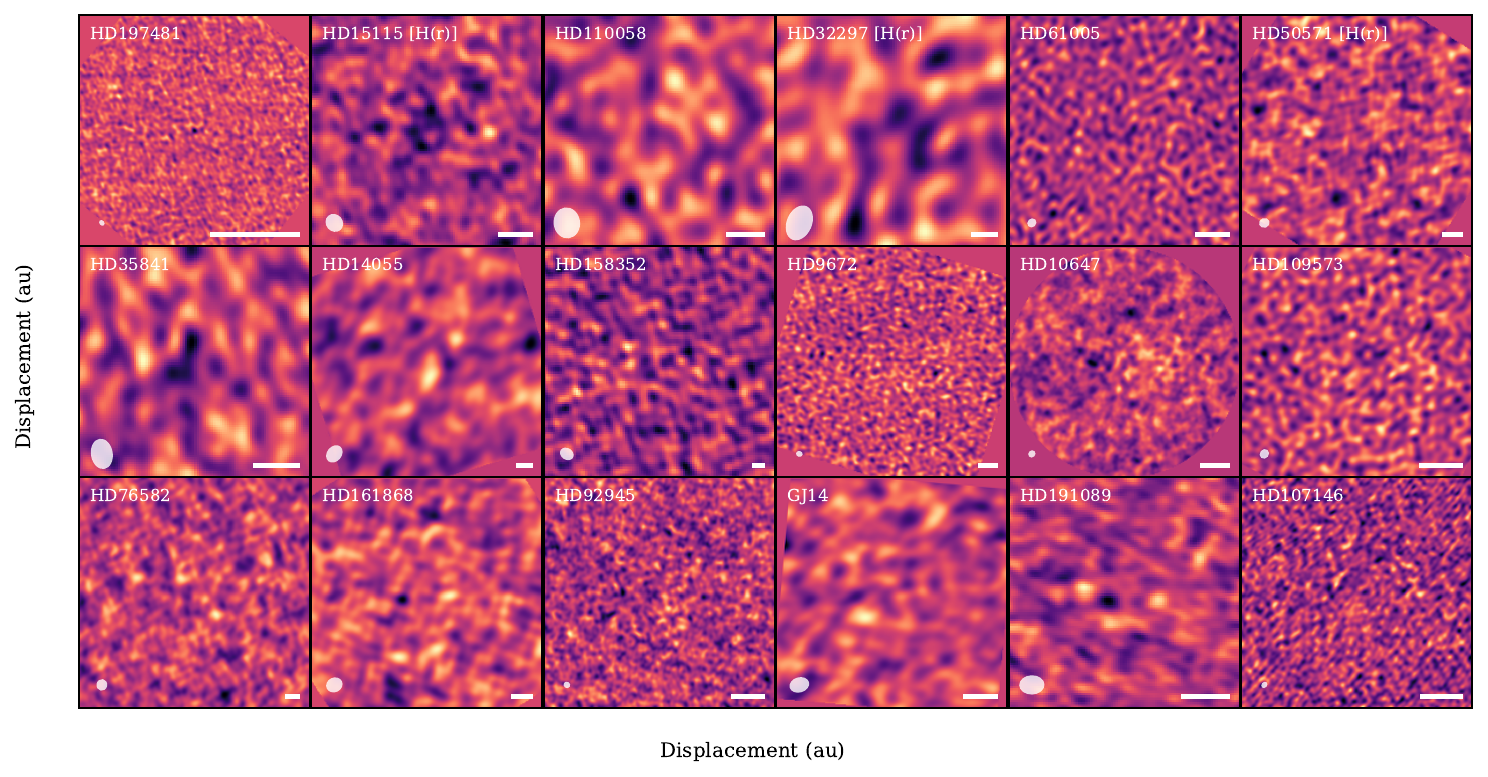}
    \caption{Residual images (data $-$ model) based on the best-fit models in Fig.~\ref{fig:model}.}
    \label{fig:2d}
\end{figure*}

\begin{figure*}
    \centering
    \includegraphics[width=18cm]{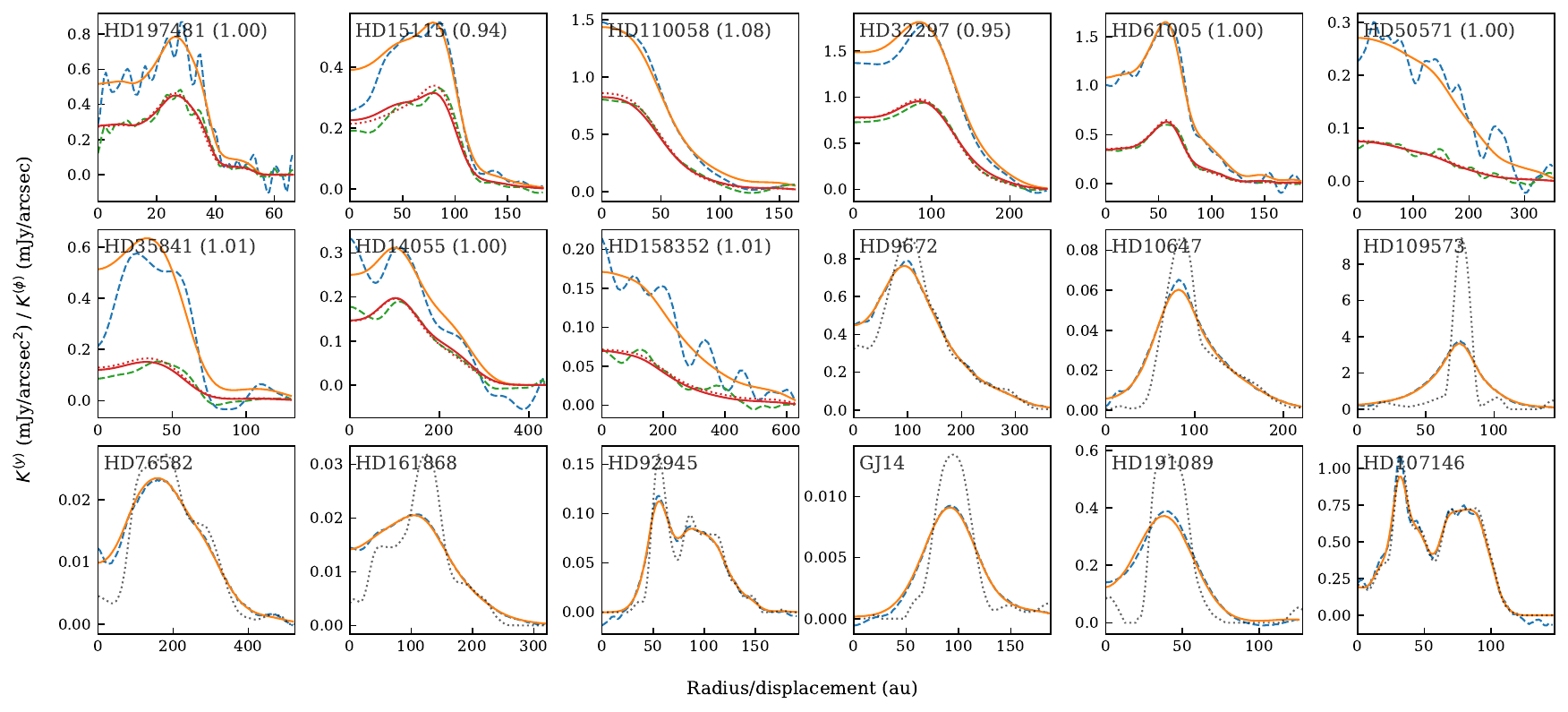}
    \caption{1D quantities of the observations and best-fit model for evaluating the goodness of fit. For the \texttt{rave} group (up to HD\,158352), the vertically summed flux ($K^{(y)}(x)$) of the data (dashed blue lines) and model (solid orange lines) and the midplane flux (within $y_\text{mid}$ from the midplane, as shown in Table~\ref{tab:targets}) of the data (dashed green lines) and model (solid red lines for the constant aspect ratio model, dotted red lines for the non-parametric height model) are plotted. The number in parentheses in each panel shows the ratio between the squared residuals of the non-parametric height profile model and the constant aspect ratio model. For the \texttt{fave} group (HD\,9672 and after), the azimuthally averaged profile ($K^{(\phi)}(r)$) of the data (dashed blue lines) and model (solid orange lines) are plotted. The radial profile de-projected and deconvolved with \texttt{fave} are shown in black dotted lines for comparison. }
    \label{fig:1d}
\end{figure*}

\bsp
\label{lastpage}
\end{document}